\documentclass{aa}

\usepackage{graphicx,verbatim,float}
\usepackage{txfonts}
\usepackage{color}

\definecolor{purple}{RGB}{76, 0,153}

\def\mpcoh{\,h^{-1}\,{\rm Mpc}}

\begin{document}

   \title{Probing the missing baryons with the Sunyaev-Zel'dovich effect from filaments}

   \author{Anna de Graaff
          \inst{1,2}
          \and
          Yan-Chuan Cai\inst{1}
          \and
          Catherine Heymans\inst{1}
          \and
          John A. Peacock\inst{1}}

   \institute{Institute for Astronomy, University of Edinburgh, Royal Observatory, Blackford Hill, Edinburgh, EH9 3HJ, UK
         \and
            Leiden Observatory, Leiden University, PO Box 9513, 2300 RA Leiden, The Netherlands \\
             \email{graaff@strw.leidenuniv.nl}
             }

   %\date{Received; accepted}

  \abstract{Observations of galaxies and galaxy clusters in the local universe can account for only $\sim\,$10\% of the total baryon content. Cosmological simulations predict that the `missing baryons' are spread throughout filamentary structures in the cosmic web, forming a low-density gas with temperatures of $10^5-10^7\,\!$K.
We search for this warm-hot intergalactic medium (WHIM) by stacking the \textit{Planck} Compton $y$-parameter map of the thermal Sunyaev-Zel'dovich (tSZ) effect for 1,002,334 pairs of CMASS galaxies from the Sloan Digital Sky Survey. 
We model the contribution from the galaxy halo pairs assuming spherical symmetry, finding a residual tSZ signal at the $2.9\mbox{$\sigma$}$ level from a stacked filament of length $10.5\mpcoh$ with a Compton parameter magnitude $y=(0.6\pm0.2)\times10^{-8}$.
We consider possible sources of contamination and conclude that bound gas in haloes may contribute only up to $20\%$ of the measured filamentary signal.
To estimate the filament gas properties we measure the gravitational lensing signal for the same sample of galaxy pairs; in combination with the tSZ signal, this yields an inferred gas density of $\rho_{\rm b}=(5.5\pm 2.9)\times\bar{\rho_{\rm b}}$ with a temperature  $T=(2.7\pm 1.7) \times 10^6\,$K. This result is consistent with the predicted WHIM properties, and overall the filamentary gas can account for $ 11\pm 7\%$ of the total baryon content of the Universe. We also see evidence that the gas filament extends beyond the galaxy pair. Averaging over this longer baseline boosts the significance of the tSZ signal and increases the associated baryon content to $ 28\pm 12\%$ of the global value.
}


   \maketitle
%
%-------------------------------------------------------------------

\section{Introduction}

The baryon content of the Universe is a key predicted observable of the standard model of cosmology. Measurements of the abundances of light elements formed in primordial nucleosynthesis and of the acoustic peaks in the power spectrum of the cosmic microwave background (CMB) indicate that baryons comprise approximately 5\% of the total energy density in the Universe \citep{Cyburt2016,PlanckCosmology}.

In the late-time Universe, however, only $\sim\,$10\% of these baryons are observed in galaxies, with another $\sim\,$10\% forming the circumgalactic and intracluster medium \citep{persic1992,Fukugita2004,shull2012}. In addition to this, \citet{McGaugh2010} reported that the baryon fraction falls below the expected universal level for all but the most massive haloes. To reconcile the apparent discrepancy in baryonic mass between the standard cosmological model and galaxy formation theories, it is therefore important to locate the remaining `missing baryons'. 

The $\Lambda$CDM model also predicts that the matter distribution of the Universe should follow a web-like pattern, in which galaxies and galaxy clusters are embedded in the knots of the web and are connected by large-scale filamentary and sheet-like structures. Both observations and simulations suggest that a significant fraction of baryons should be found outside the gravitationally bound haloes, in filaments and sheets \citep{hydrosim1999,penton2004,shull2012}. The baryons are expected to be in a diffuse `warm-hot' state, with low over-densities ($\delta\sim10$) and temperatures between $10^5-10^7\,\!$K. 

Previous efforts to detect this warm-hot intergalactic medium (WHIM) focused mainly on the measurement of absorption lines in the spectra of distant quasars. 
\citet{penton2004} measured the Lyman-$\alpha$ forest at low redshift and estimated a baryon density of $0.3\Omega_{\rm b}$. The Lyman-$\alpha$ absorbers trace relatively cool gas ($T\sim 10^4\,$K), but absorption lines of highly ionised heavy elements have also been observed, requiring higher temperature gas  \citep[e.g.][]{Nicastro2008,Tejos2016}. Combined, these absorption line studies can account for $\sim\,$50\% of baryons. Most recently, \citet{Nicastro2018} reported the detection of the WHIM from the measurement of two OVII absorbers, and extrapolate that their measurement may account for all of the missing baryons, but with large uncertainties.

A second successful approach to the detection of the diffuse WHIM was the individual detection of filaments using their X-ray emission \citep{kull1999,xray:filaments}. These X-ray studies however probe mainly the high-temperature ($T\sim10^7\,$K) and high-density ($\delta\gtrsim 100$) end of the WHIM, and thus only yield detections near massive clusters. 

In summary, these different methods are sensitive mostly to the lower and higher temperature end of the warm-hot baryons, leaving the majority of the missing baryons still poorly constrained \citep{shull2012}. In particular, the observations of individual lines of sight or single objects cover a small volume of the Universe, suffering from large cosmic variance and giving no clear picture of the large-scale extent and geometry of the WHIM.

The Sunyaev-Zel'dovich (SZ) effect provides an alternative means of detecting the WHIM in filaments. The thermal SZ effect arises from the Compton scattering of CMB photons by ionised gas. The amplitude of the signal is quantified by the Compton $y$-parameter \citep{SZ1972}, 
\begin{equation}
y = \int\frac{k_\mathrm{B} T_\mathrm{e}}{m_\mathrm{e} c^2} \, n_\mathrm{e} \, \sigma_\mathrm{T} \, \mathrm{d}\ell,
\label{eq:tsz}
\end{equation}
where $m_\mathrm{e} c^2$, $k_\mathrm{B}$ and $\sigma_\mathrm{T}$ are the electron rest mass energy, Boltzmann constant and the Thomson cross-section respectively, all of which are known physical quantities that amount to a constant. The $y$-parameter therefore depends solely on the line-of-sight integration of $n_\mathrm{e} T_\mathrm{e}$, which are respectively the electron number density and temperature. The kinetic SZ effect on the other hand is sensitive to the bulk momentum of free electrons in and around clusters and galaxies.

Both the thermal and kinetic SZ effect have been observed with high significance for the high-density environments of galaxies, galaxy clusters and their outskirts (e.g. \citealt{kSZ2015HM, Hill2016, Schaan2016, KSZ_Plannck2016,DeBernardis2017}; for a recent review see \citealt{Mroczkowski2018}). Notably, \citet{Bonjean2017} and \citet{PlanckFilament} report the tSZ observation of a bridge connecting the merging cluster pair A399-A401, and conclude this constitutes a detection of the hottest and densest part of the WHIM. 

In this work we search for the fainter predicted thermal SZ signal from large-scale gas filaments using the all-sky Compton parameter $y$-map from \citet{planck2016ymaps}; because the imprint of the WHIM signal on the $y$-map is expected to be weak, we proceed by stacking, so the locations and orientations of filaments on the map need to be known in advance. 
From analyses of numerical simulations, we expect filaments to connect pairs of massive haloes separated by up to $\sim 20\mpcoh$ \citep{colberg2005}. As such, a good proxy for the location of filaments is the line connecting neighbouring massive galaxies \citep{Clampitt2016,hudson2017}.

In section~\ref{sec:data} we describe our galaxy pair catalogue and the \textit{Planck} data used. Our stacking analysis is presented in section~\ref{sec:stacking}, followed by a discussion of possible contaminants and the estimation of the filament properties in section~\ref{sec:interpretation}. In sections~\ref{sec:discussion}~\&~\ref{sec:conclusion} we compare our result with current literature, including the recent similar SZ stacking analysis by \citet{tanimura2017}, and present our conclusions.

Throughout the paper we assume a flat $\Lambda$CDM cosmology with Hubble constant $ H_{0}= 100\,h\rm\,km\,s^{-1}\,Mpc^{-1}$, $h=0.68$, $\Omega_{\rm m}=0.31$ and $\Omega_{\rm b}=0.049$ \citep{PlanckCosmology}.

\section{Data}\label{sec:data}

\subsection{Galaxy pair catalogues}
\subsubsection{Filament candidates}

We use both the North and South CMASS galaxy catalogues from the 12\textsuperscript{th} data release of the Sloan Digital Sky Survey (SDSS: \citealt{sdss:dr12,boss:dr12}). The CMASS galaxies were selected using colour-magnitude cuts to identify galaxies in the redshift range $0.43<z<0.75$ with a narrow range in stellar mass. The galaxies have a mean stellar mass of $10^{11.3}M_\odot$ and are mostly central galaxies in their host dark matter haloes of typical virial mass $\sim10^{13}\,h^{-1}\,M_\odot$ \citep{Maraston2013,White2011}. CMASS galaxies therefore represent a highly biased galaxy sample. 

Using the full CMASS catalogue of 0.85 million galaxies, we select a sample of galaxy pairs that are likely to be connected by filaments (`physical pairs'). Motivated by \citet{Clampitt2016}, the pairs are required to have a transverse comoving separation in the range $6-14\mpcoh$ and a line-of-sight separation $<5\mpcoh$\footnote{Our criterion on line-of-sight separation implicitly assumes that the total redshift differences are purely cosmological; but in reality, peculiar velocities can have an effect — as can be seen from e.g. Fig.~11 of \citet{Jenkins1998}. For filaments of our length that are near to the plane of the sky, the pairwise dispersion in radial velocity is close to 500~km/s, and this has a convolving effect. Thus, some of our pairs will have a true radial separation slightly larger than our limit of $5\mpcoh$, and some pairs with a smaller radial separation will be rejected because the apparent separation in redshift space is above our limit. But our selection still picks pairs that are nearly transverse to the line of sight, and is distinct from the much larger radial offset that we use elsewhere to select control samples of unphysical pairs for control purposes.}. 
These ranges were chosen to ensure that the intergalactic medium is probed as well as the intracluster medium within the haloes (of virial radius $\sim1\mpcoh$). By applying the first constraint we prevent contamination from the potential projection of two haloes along the line of sight. The final selection of physical pairs comprises 1,020,334 pairs with a mean angular separation of 26.5 arcmin and a mean comoving separation of $10.5\mpcoh$.
The number of galaxy pairs exceeds the number of galaxies, indicating that some of the CMASS galaxy haloes may be undergoing collapse into galaxy groups or clusters. As a result, a small fraction of the CMASS galaxies will overlap with the expected filament region. We account for this effect when estimating the significance of the SZ signal in section~\ref{sec:covariance} and estimate the contribution from CMASS galaxy haloes to the filament signal in section~\ref{sec:gasinhaloes}.

\subsubsection{Control sample}\label{sec:nonphys}

In addition to our filament candidates, we compile a second sample of `non-physical' pairs of galaxies that have the same comoving projected separation of $6-14\mpcoh$, but are separated by $40-200\mpcoh$ along the line of sight.
These pairs of galaxies therefore have the same projected separation as our sample of physical pairs, yet are highly unlikely to be connected by filaments \citep{hudson2017}. The resulting selection of 13,622,456 non-physical pairs has a distribution of angular separations similar to that of the physical pairs. 
We use this catalogue to estimate the contribution to the SZ signal from uncorrelated large-scale structures that lie along the line of sight to the galaxy pairs.

\subsection{\textit{Planck} maps}
\subsubsection{Compton parameter map}

We use the MILCA (Modified Internal Linear Combination Algorithm: \citealt{hurier2013}) and NILC (Needlet Internal Linear Combination) all-sky Compton parameter maps (`$y$-maps') released by the Planck Collaboration \citep{planck2016ymaps}. These $y$-maps were constructed from the multiple \textit{Planck} frequency channel maps, 
which were convolved to yield a common circular Gaussian beam of FWHM 10 arcmin \citep{planck2016ymaps}. We apply a $40\%$ Galactic mask provided by the Planck Collaboration to reduce contamination from galactic emission and point sources. Both the $y$-maps and mask are provided in the {\sc HEALPix} \citep{gorski2005} format at a resolution of $N_\mathrm{side}=2048$. Here we present only our results obtained with the MILCA $y$-map, however we note that our resuls are unchanged if we replace the MILCA $y$-map with the NILC $y$-map in our analysis.

\subsubsection{CMB lensing convergence map}\label{sec:lensingmap}

Large-scale matter over-densities will cause coherent distortions of the CMB temperature fluctuations \citep{Plancklensing}. This lensing effect can then be used to measure the dimensionless projected matter density along the line of sight, the convergence $\kappa$: 
\begin{equation}
\label{kappa}
\kappa({\bf r})=\frac{3H_0^2\Omega_{\rm m}}{2c^2} \int_{0}^{D_\mathrm{S}}\frac{(D_\mathrm{S}-D_\mathrm{L}) D_\mathrm{L}}{D_\mathrm{S} }\,\frac{\delta({\bf r}, D_\mathrm{L})}{a} \, dD_\mathrm{L}; 
\end{equation}
here, {\bf r} is the line-of-sight direction, $c$ is the speed of light, $D_\mathrm{L}$ and $D_\mathrm{S}$ are the comoving distances of the lens and the source, which is the distance to the last scattering surface for the case of CMB lensing.  $\delta \equiv \rho/\bar{\rho}-1$ is the 3D density contrast, with $\rho$ being the total local matter density and $\bar{\rho}  = 3\Omega_{\rm m}H^2/(8\pi G) $ the mean matter density of the Universe, where $H$ is the Hubble parameter and $G$ the gravitational constant.

In addition to the SZ $y$-map, we use the CMB lensing convergence map ($\kappa$-map) from \textit{Planck} \citep{Plancklensing} to estimate the matter density around the galaxy pairs (see section~\ref{sec:gas_props}). 
We convolve the $\kappa$-map with a Gaussian beam of FWHM 10 arcmin to make sure that it has the same small-scale coherence as the $y$-map
(the $\kappa$-map is a reconstruction that does not possess a beam: only a harmonic limit of $\ell<2048$).

\section{Stacking analysis}\label{sec:stacking}

\subsection{Compton parameter map stacking}

As the expected SZ signal from a single filament falls well below the noise level of the $y$-map, we stack the signal from the full sample of galaxy pairs.
The map area surrounding each galaxy pair is rotated such that the galaxy pair aligns with the equator, with the centre of the pair located at the origin of the galactic coordinate system. The rotated areas are then rescaled according to their corresponding angular pair separation, such that the locations of the different pairs overlap with each other. We then project each map onto a two-dimensional rectangular grid using a nearest neighbour interpolation. 
For every galaxy pair we stack both the projected map and its mirrored version, thus resulting in a stacked map symmetrical in the vertical axis. Masked {\sc HEALPix} pixels are accounted for by assigning a weight of zero to the corresponding grid pixels.

\begin{figure*}
\centering
\includegraphics[width=\linewidth]{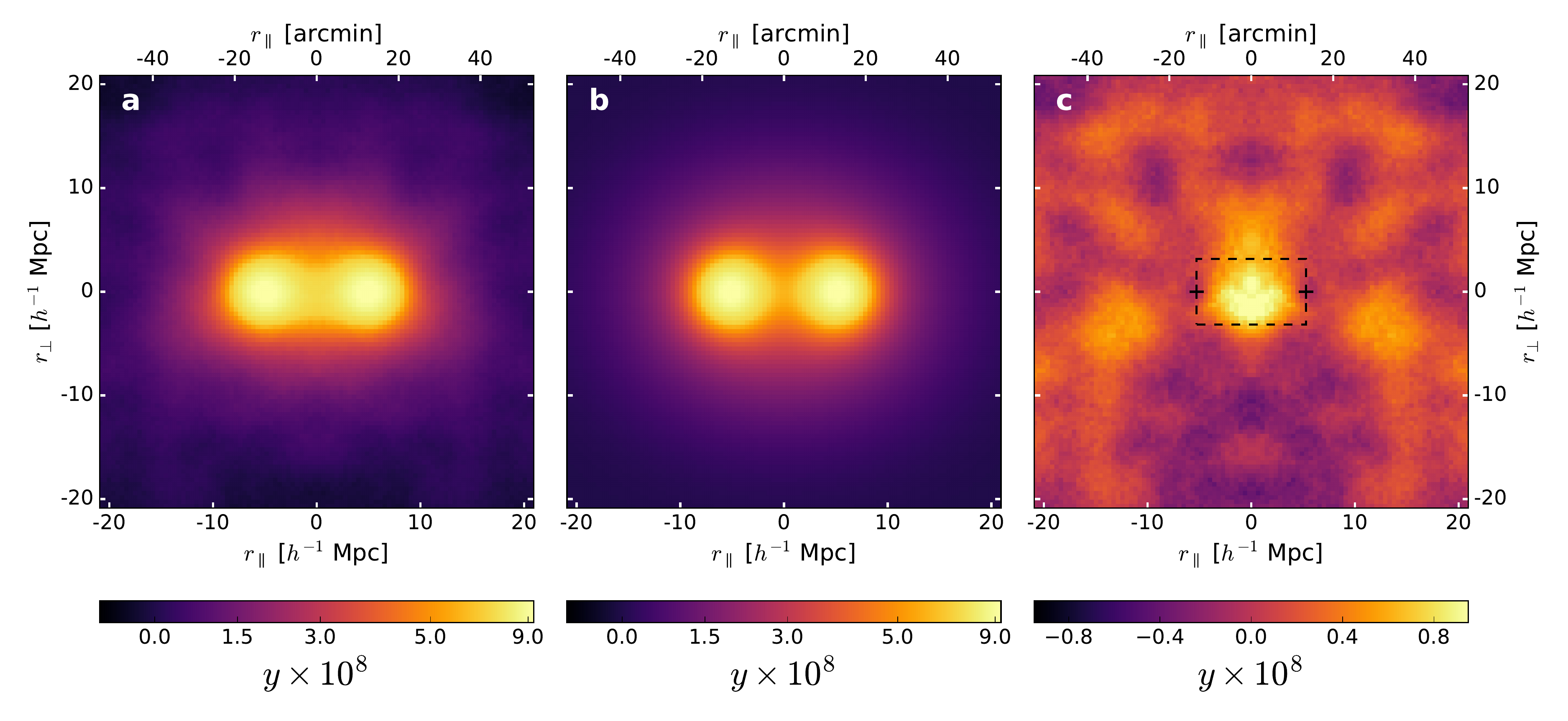}
\caption{(a) Symmetrically stacked Compton $y$-parameter maps for 1 million close pairs of CMASS galaxies; (b) the modelled signal from the galaxy host haloes only; and (c) the residual between the stacked data and model.
The indicated horizontal and vertical distance scales ($r_\parallel$ and $r_\bot$ respectively) are calibrated using the mean galaxy pair separation of $10.5\mpcoh$. The mean projected angular separations are also shown for the horizontal axis. There is a bridge connecting the pairs of galaxies in the data (a) but not in the model (b), suggesting the presence of a filament in (a), which is highlighted in panel (c) by a dashed box with plus signs indicating the positions of the galaxy pairs.}
\label{fig:2D_plots}
\end{figure*}

In order to improve the computational efficiency of our stacking algorithm, we reduce the map resolution to $N_\mathrm{side}=1024$, corresponding to a typical pixel size of 3.4 arcmin.  
The resulting stacked $y$-map from the 1 million pairs is presented in the left panel of Fig.~\ref{fig:2D_plots}. The map is dominated by the SZ signal from the two bright galaxies; but the main important feature is the bridge of emission that connects the pair, which suggests the presence of a filament. 

\subsection{Modelling the galaxy haloes}

\begin{figure}
\centering
\includegraphics[width=\linewidth]{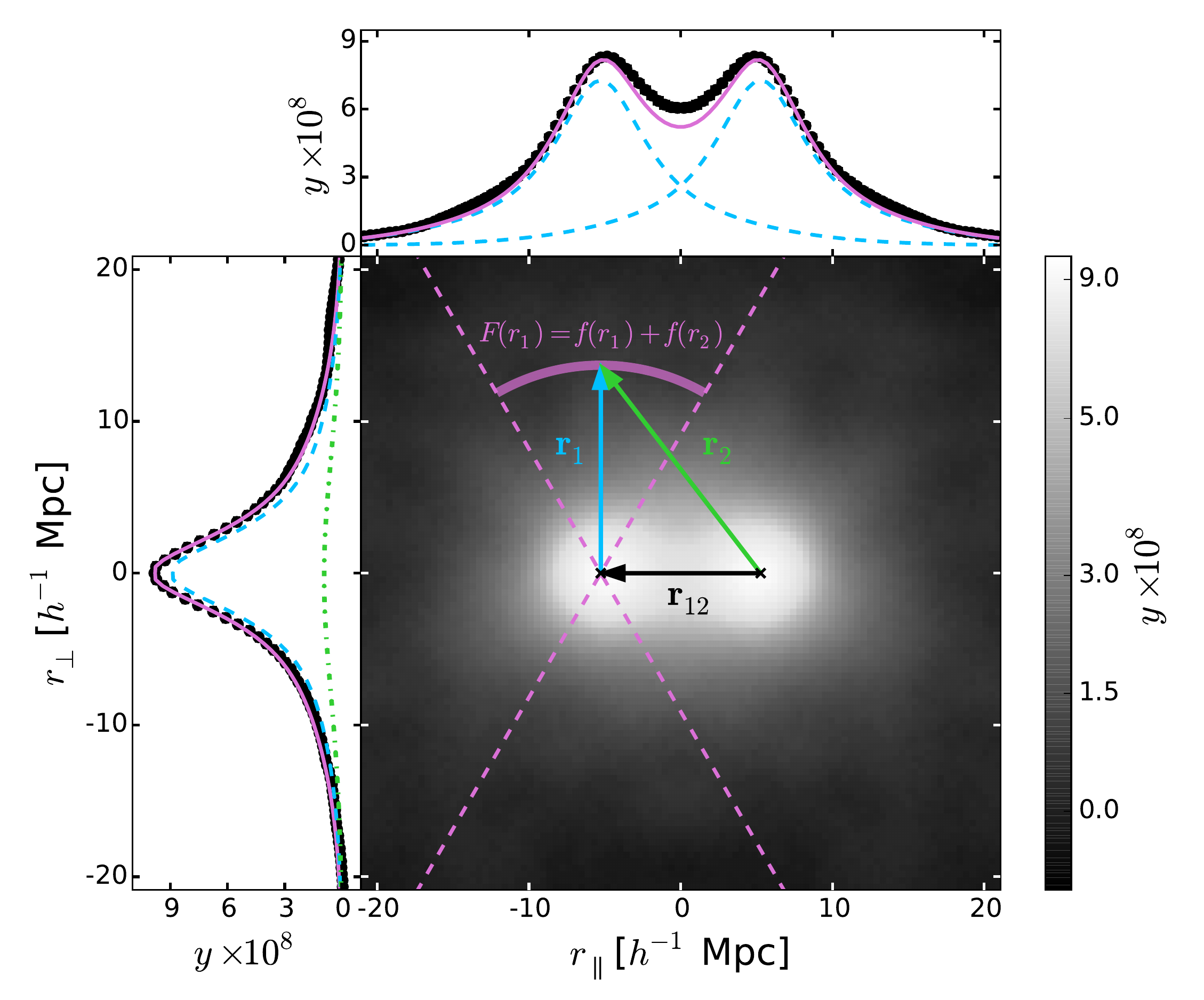}
\caption{Illustration of the fitting procedure used to decompose the contribution from the isotropic haloes and the filament. The stacked CMASS galaxy pairs for the Compton $y$-map are shown in black and white in the main panel. The mean horizontal profile extracted from the 2D plot is in the upper panel. The pink dashed lines indicate the 60-degree subtended angle used to construct the mean radial profile in the left panel. The arrows demonstrate how the two haloes were decomposed for the halo modelling. Blue colours correspond to the primary halo contribution, green to the secondary halo, and pink to the combined contribution from the two haloes. $F(r)$ indicates 
the sum of the two isotropic halo profiles [$f(r_1)$ and $f(r_2)$] along the vertical direction.}
\label{fig:diagram}
\end{figure}

\begin{figure*}
\centering
\includegraphics[width=1.0 \linewidth]{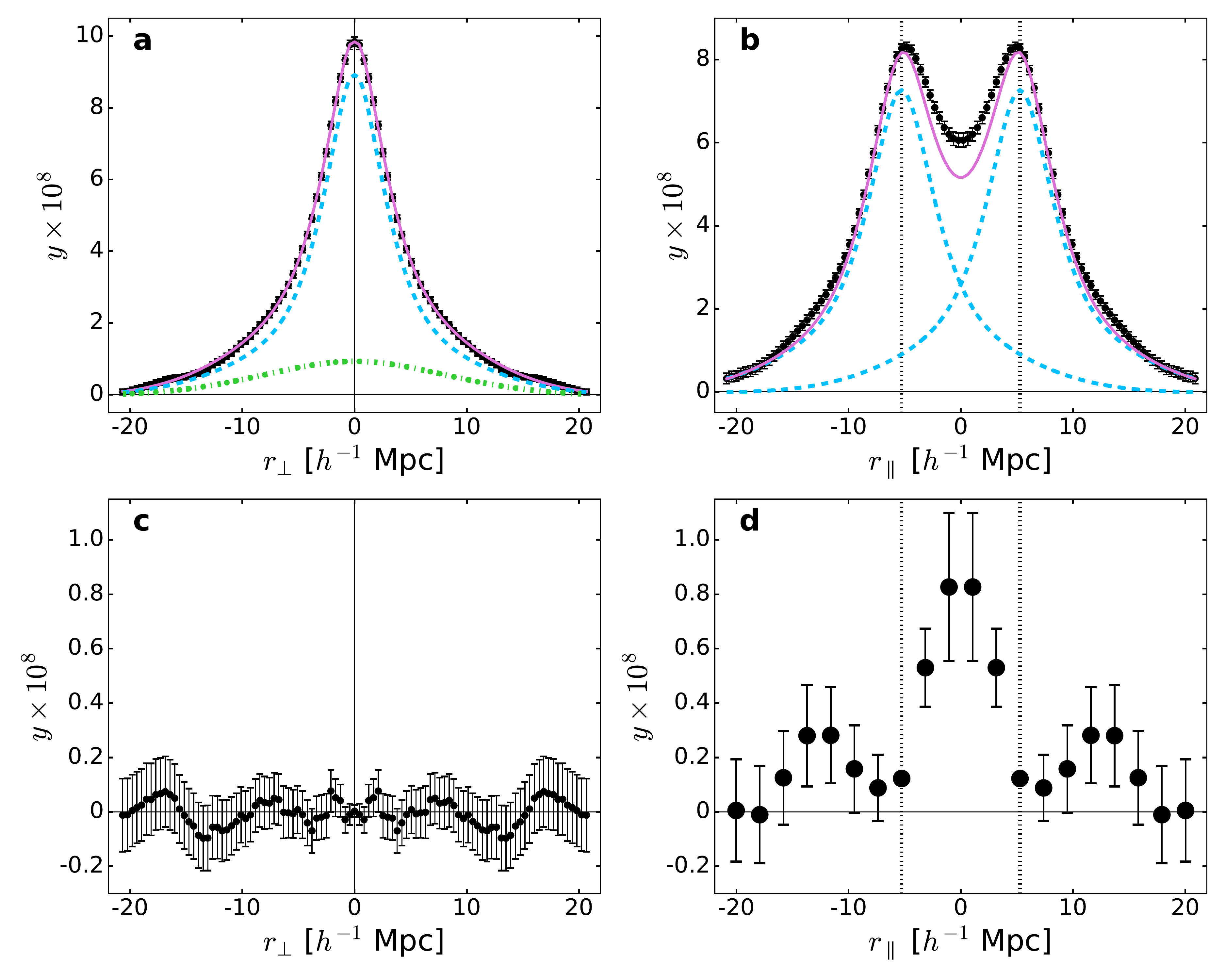}
\caption{One-dimensional profiles of the SZ signal from stacked galaxy pairs. (a) The mean radial SZ profile extracted along the vertical direction in the left-hand panel of Fig.~\ref{fig:2D_plots}; (b) the mean horizontal profile with a thickness of $6\mpcoh$ extracted from the left-hand panel of Fig.~\ref{fig:2D_plots}; the residual of (c) the radial profile and (d) the horizontal profile after subtracting the contribution from the two haloes. Error bars in panels a and b represent simple estimates of the errors in the mean profile, while in panels c and d the error bars are obtained from a detailed jackknife analysis. The blue dashed lines and the green dash-dotted line indicate the modelled primary and secondary halo contributions respectively. The pink solid lines show the combined modelled contribution from the two haloes. The residual in (c) is consistent with zero, indicating the success of our modelling. The detected filament lies interior to the halo centres  (dotted lines), shown by the offset between the solid pink line and the black data points in (b), and by the residuals in (d).} 
\label{fig:1D_filament}
\end{figure*}

Both galaxies in the pair will be surrounded by
an isotropic `SZ halo' signal, arising from the gas in their
parent haloes, plus gas in other correlated haloes, all
convolved with the extended \textit{Planck} beam.
We model the overall stacked SZ map (Fig.~\ref{fig:2D_plots}) as a superposition of two such isotropic profiles, in order to look for a residual filament signal along the line between the pair. 
The isotropic fitting is performed using only the data within a 60-degree subtended angle along the vertical direction (as indicated in Fig.~\ref{fig:diagram}). This vertical selection is chosen to exclude possible contamination from a filament, which is most likely to be along the horizontal direction.
We assume the radial $y$-profile of the halo can be described by some unknown function $f(r)$. Since the stacking procedure is symmetrical about the vertical axis, the two haloes are described by the same function. Along the vertical direction, at a radius $r_1$ from the halo centre on the left, the SZ signal contains contributions from both haloes. The total SZ signal is then $F(r_1)$, where $F(r_1)=f(r_1)+f(r_2)$ and $r_2=(r_1^2+r^2_{12})^{1/2}$ (see Fig.~\ref{fig:diagram}). 
We use a fourth order polynomial multiplied by an exponential function for $f(r)$, which is found to provide an accurate fit to the data with residuals at the $1\times 10^{-9}$ level (left-hand panels of Fig.~\ref{fig:1D_filament}). From this model we then generate a 2D $y$-map of the two haloes, which is shown in the middle panel of  Fig.~\ref{fig:2D_plots}. The difference between this model and the stacked data (left) is shown on the right. No significant residual of the two main haloes is found, with the exception of the region interior to the halo centres (indicated by the dashed box), as is expected if there is an SZ signal associated with filaments. There is a visible asymmetry along the vertical direction in Fig.~\ref{fig:2D_plots}. We have checked that only a minor part of this ($\sim$ 10\%) is caused by the projection of {\sc HEALPix} pixels onto the Cartesian grid. The asymmetry is mostly random and consistent with sample variance, which we have confirmed using the sample of non-physical pairs. The extracted one-dimensional horizontal profiles for the stacked data and model are presented in the upper right panel of Fig.~\ref{fig:1D_filament}.
The corresponding one-dimensional residual signal is shown in the lower right panel.

The filament signal is at most $\sim\,$10\% of the overlapping
isotropic signals from the two haloes, so some care is
needed in order to be convinced that the residual cannot
be an artefact of any error in the assumed halo profiles. In
particular, if the true halo profiles were slightly broader than our best-fit model, this would raise the signal in the overlap region between the haloes. To investigate this, we introduced an additional nuisance parameter, in which the true halo profile is expanded in radius by a factor $s$ compared to our estimate. The parameter $s$ was then allowed to float in order to best-fit the full 2D data
in Fig.~\ref{fig:2D_plots}, adopting the null hypothesis that no filament is present. We also allowed a free vertical normalisation in this exercise. In fact, the preference is for a slightly  narrower profile, with $s=0.94$, driven by the negative residuals at large $r_\perp$ and small $r_\parallel$. In any case, this scaling changes the inferred filament signal by only $\sim\,$10\%. We therefore conclude that our main result is robust with respect to the assumed model for the isotropic halo profile.

\subsection{Estimating the significance of the filament signal}\label{sec:covariance}

To estimate the significance of the tentative filament signal, we measure the profile of the residual $y$-map along the horizontal direction with a width of $\sim\,$$6\mpcoh$, as indicated by the right-hand panel in Fig.~\ref{fig:2D_plots}. 
The width of the filament was chosen to be approximately 1.5$\times$FWHM of the beam profile for the $y$-map.
We first construct the covariance of the data points using individual $y$-maps associated with each pair,
\begin{equation}
C_{i,j}=\frac{1}{N}\sum_{k=1}^N(y^k_i-\bar{y_i})(y^k_j-\bar{y_j}),
\end{equation}
where the subscript $i$ and $j$ indicate the indices of the bins, the superscript $k$ represents the index of galaxy pair and $N=1,020,334$ is the total number of pairs.  
This is an estimate of the covariance of a single map in the stack, but we want the covariance of the mean. If all maps were independent, this would require
$C\to C/N$. We start by
computing the total $\chi^2$ value on this assumption:
\begin{equation}
\chi^2=\sum_{i,j}^n \bar{y_i} C^{-1}_{i,j} \bar{y_j},
\label{eq2}
\end{equation}
where $n$ is the number of bins. The $\chi^2$ values were converted into the corresponding Gaussian $\sigma$ values taking into account the number of degrees of freedom. 
The data points between the two galaxies, which we interpret as being due to gas filaments, deviate from zero at the $5.1\sigma$ confidence level. We also find a lower-significance horizontal excess of the SZ signal outside the galaxy pairs, which may be the extension of the filament.

{However, treating all galaxy pairs as giving independent estimates of the SZ signal will not be correct if the filament regions for each galaxy pair overlap with each other. The DR12 CMASS sky area is 9376\,deg$^2$, whereas the rectangle assumed to enclose the filament is approximately 0.12\,deg$^2$. Thus with $10^6$ filaments, each SZ pixel in the CMASS region must appear in the stack approximately 13 times and we will underestimate the errors if we assume all maps in the stack to be independent. But we cannot simply scale all errors by $\sqrt{13}$ as the fraction of duplicated pixels will depend on position within the map. The only feasible routes for obtaining a correct covariance matrix on the stack are then either to average over a large ensemble of realistic simulations, or to use a resampling strategy. We choose the latter approach, and present below detailed results using the 
jackknife method to estimate the significance of the signal.}

We split the galaxy pairs into sub-samples $N_{\rm sub}$ according to their sky coordinates, such that each sub-sample is of approximately equal area and number of galaxy pairs. For $N_{\rm sub}=250$ this corresponds to regions of $\sim\,40\,$deg$^2$ in size containing on average $N'_{\rm pair}\simeq 4000$ pairs. Since the average angular separation of the galaxy pairs is $0.44\,$deg, these regions are sufficiently large to justify treating the areas spanned by the sub-samples as independent. 
Each jackknife sub-sample then contains $N_{\rm pair}=N-N'_{\rm pair}$ pairs of galaxies. The modelling analysis is then repeated for the $N_{\rm sub}$ jackknife sub-samples, based on the mean measurements of the $N_{\rm pair}$ pairs in each jackknife sub-sample. We use the resulting $N_{\rm sub}$ residual profiles ($y$) to construct the covariance,

\begin{equation}
C^{\rm JK}_{i,j}=\frac{N_{\rm sub}-1}{N_{\rm sub}} \sum_{k=1}^{N_{\rm sub}}(y^k_i-\bar{y_i})(y^k_j-\bar{y_j}),
\end{equation}
where $i$ and $j$ still represent the indices of the bins, but $k$ now indicates the index of the sub-sample and
\begin{equation}
\bar{y_i} = \sum^{N_{\rm sub}}_{k=1} y^k_i/N_{\rm sub}.    
\end{equation} 

{As before (Equation~\ref{eq2}) we compute the $\chi^2$ value to compute the significance, however we now also add a correction factor to account for bias in the inverse covariance matrix \citep{Hartlap2007}, such that
\begin{equation}
    \chi^2=\sum_{i,j}^n \bar{y_i} \left ( \frac{N_{\rm sub}-n-2}{N_{\rm sub}-1} \right ) \left [ C^{\rm JK}_{i,j} \right]^{-1} \bar{y_j},
\end{equation}
where $n$ again indicates the number of bins.
For $N_{\rm sub}=250$ we find the significance of the filament ($n=12$) to be $2.9\sigma$. 
To test this result, we have repeated the jackknife analysis for different values of $N_{\rm sub}$ (ranging from $N_{\rm sub}=100$ to $N_{\rm sub}=1000$), finding convergence to within $\pm 0.3\sigma$ from the quoted $2.9\sigma$ result. Including the data points outside the galaxy pairs increases the significance to $3.8\sigma$; for the calculation of this significance we have excluded the data point at the location of the galaxy centres  as a precaution, since the jackknife error here is very small ($2\times10^{-10}$) and we want to avoid possible contamination from the galaxy to the filament.}

We note that, in order to perform the model fitting for the mean vertical profile of each jackknife region, we require error estimates of the vertical profile. These are initially taken to be the simple independent-snapshot estimate of the errors on the mean, as of course we start without a knowledge of the jackknife errors. But this is not a critical issue: the errors on the profile are very small and do not dictate the uncertainty in the residual. Also, the true errors resulting from this fitting process are by construction correctly represented by the resampling. We experimented with ignoring errors entirely and fitting the profiles by least squares; this changed the significance of the filament residual by only $+0.1\sigma$, so we conclude that profile fitting uncertainties are sub-dominant to the main sources of error discussed in the following section.

\section{Interpretation of the filament signal}\label{sec:interpretation}

Having identified a filament signal in the stacked $y$-map, we now need to ask whether this constitutes a detection of the WHIM, or whether the signal might have some other origin. There are three possible sources of contamination that must be considered: (i) uncorrelated sources, including foreground and background dusty galaxies, and dust emission from the Milky Way; (ii) correlated SZ emission from gas in galaxy haloes that follow the filament between our galaxy pairs; (iii) systematic signal leakage from the Cosmic Infrared Background (CIB) into the \textit{Planck} $y$-map. 

\subsection{Uncorrelated sources}

We estimate the residual SZ signal due to the uncorrelated large-scale structures by repeating our stacking and fitting procedure for the catalogue of non-physical galaxy pairs (section~\ref{sec:nonphys}). We use 13.6 million selected non-physical CMASS galaxy pairs to draw 500 subsamples of equal size to the sample of physical pairs, and then perform the stacking, halo modelling and profile extraction for each subsample.
We find the mean residual SZ signal for the non-physical pairs to be consistent with zero, and therefore conclude that uncorrelated large-scale structures, including dust emission from the Milky Way, cannot make a significant contribution to the SZ signal of the detected filament.

\subsection{Bound gas in correlated haloes}\label{sec:gasinhaloes}

\begin{figure*}[p]
\centering
\includegraphics[width=0.9\linewidth]{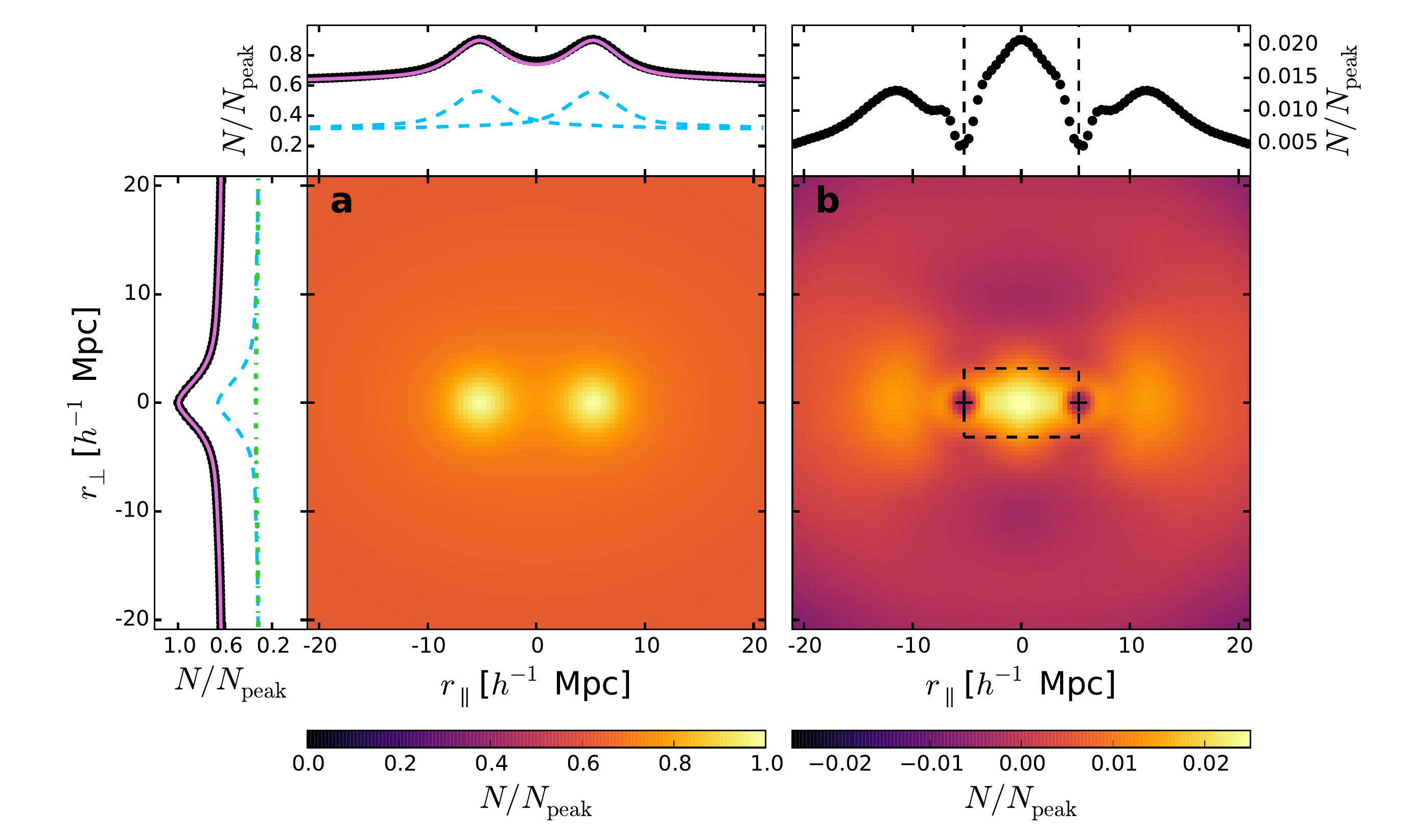}
\caption{(a) Main panel: the stacked CMASS galaxy number density map for 1 million galaxy pairs, convolved with a Gaussian filter with FWHM 10 arcmin and normalised by the peak value of the stacked map. The left and upper side-panels show the mean radial profile and horizontal profile respectively, and include the modelled contributions from the primary halo (blue dashed lines), secondary (green dash-dotted line) and the two haloes combined (pink solid lines). (b) The residual between the stacked map and two isotropic halo profiles. The upper side-panel shows the residual profile extracted from the boxed region, with dashed lines indicating the galaxy pair centres. 
The residual haloes in the filament region are estimated to contribute approximately 20\% of the filament signal.}
\label{fig:num_density}
\end{figure*}

\begin{figure*}[p]
\centering
\includegraphics[width=0.9\linewidth]{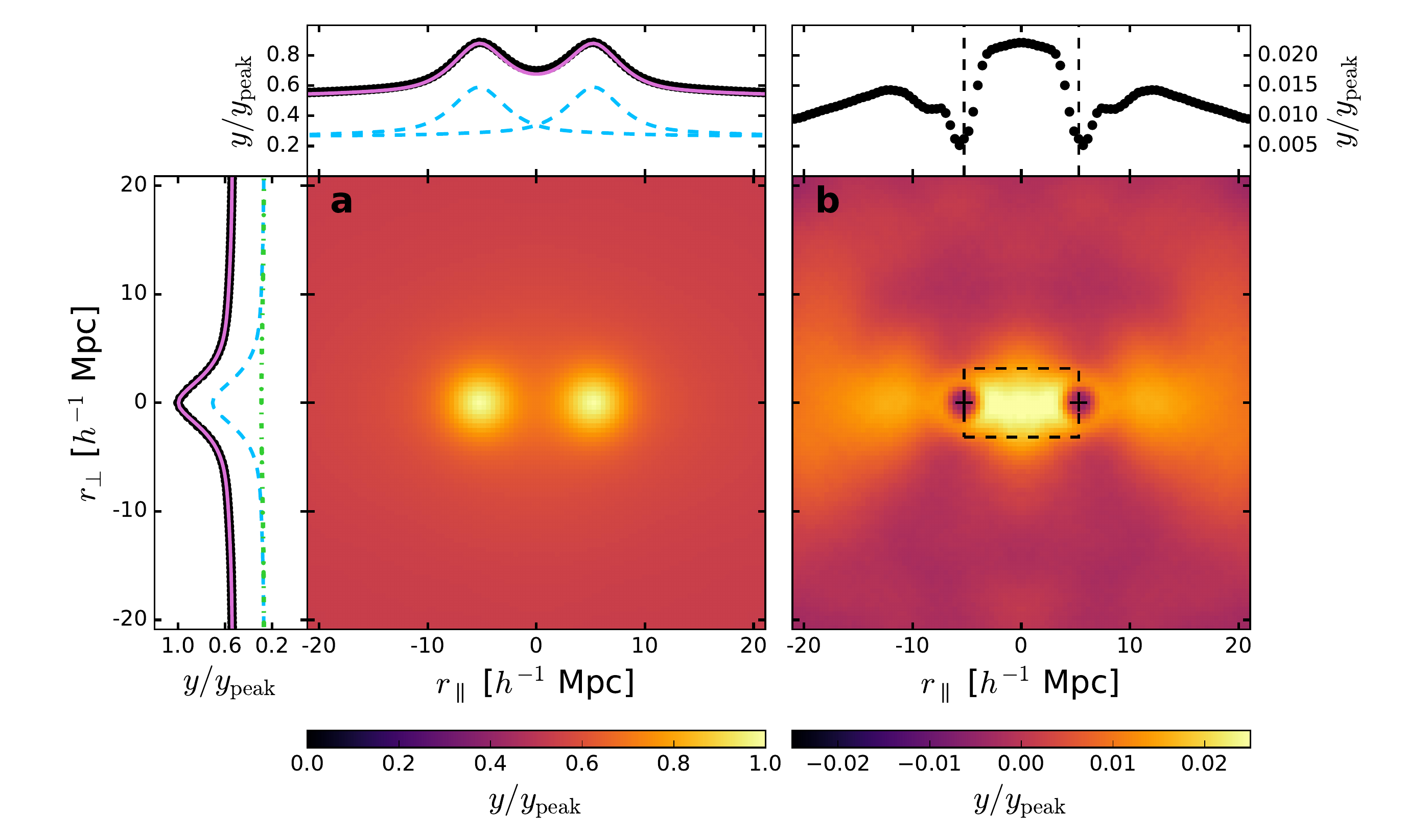}
\caption{Panels show the same as in Fig.~\ref{fig:num_density}, except that the colour bars here represent the normalised $y$-value from our simulation.
These results from simulation show the contribution to the filament from gas haloes, rather than diffuse gas. The residual haloes in the filament region are estimated to contribute approximately 20\% of the filament SZ signal, consistent with the results obtained from Fig.~\ref{fig:num_density}.}
\label{fig:Millennium_2D_plots}
\end{figure*}

The non-physical pairs lack the contribution to the SZ signal that can arise from haloes that lie along the filament between the galaxy pairs. We examine the effect of these possible correlated structures using both observations and simulations. 

We start by constructing a map of the galaxy number density for the whole CMASS sample. 
The map was convolved with a Gaussian filter of FWHM 10 arcmin to represent the beam size of the $y$-map, and we then apply our stacking and fitting procedure to this map. The resulting 2D stacked pairs and residual profiles are shown in Fig.~\ref{fig:num_density} together with the extracted profiles and fitted models. After subtraction of the isotropic component of the two haloes, we find that a filament in the light distribution between the galaxy pairs is detected. However, the relative level of this signal is small: 1-2\% of the peak values of the two haloes, as against $\sim\,$10\% in the SZ analysis. From this we can then estimate that CMASS galaxy haloes within the filaments contribute less than 20\% of the detected filament signal.

But in making the estimate of the total contribution to the filament signal from galaxies within the filament, we should also consider the contribution from galaxies below the CMASS limit. These are significant in principle: taking the relation between the SZ decrement and stellar mass from \citet{Greco2015} and the stellar mass function at $z=0.55$ from \citet{Maraston2013}, we estimate that galaxies at or above the CMASS stellar mass contribute approximately 1/3 of the global SZ signal. However, if the low mass galaxies cluster about CMASS galaxies in the same way as CMASS galaxies cluster together, then their contribution would not change our analysis: they would enhance the signal near the CMASS galaxies and along the filament, so that their contribution would scale out.
Because of the rapid scaling of SZ decrement with galaxy mass, the most important missing haloes are only a few times smaller than the CMASS haloes, so any difference in their clustering bias parameter would be at the level of 10\%, suggesting that they would make a sub-dominant change to the 20\% SZ contribution that we estimated above.

A caveat of the above estimate is that the galaxy number density map is only an approximation for the $y$-map intensity contributed by CMASS galaxies. The $Y-M$ relation, where $Y$ is the total SZ Comptonisation parameter within a certain radius and $M$ is the halo mass enclosed within the same radius, is effectively taken as $Y\propto M^{\beta}$ with the slope parameter $\beta \simeq 1$. This is expected from halo occupation distribution (HOD) fitting results for the CMASS galaxy sample \citep{Manera2013, White2011}. However, $\beta$ is predicted to be $5/3$ from the self-similar model \citep{daSilva2004,Motl2005, Bonamente2008,Kay2012, Sembolini2013, planck2016ymaps}. We therefore verify our estimate of the contribution to the excess SZ signal from haloes outside of the galaxy pairs using an independent estimate from simulations in which the $Y-M$ relation has been accounted for directly.

We construct a full-sky $y$-map using the Millennium N-body simulation \citep{Springel2005}. We use all haloes above the mass of $10^{10}\,h^{-1}M_{\odot}$, as these host most of the galaxies that may have gas haloes contributing to the SZ signal of the filament. Each halo is assigned a $y$-value according to the well studied $Y-M$ relation with slope $\beta=5/3$. 
Haloes in the Millennium simulation box at $z\simeq 0.5$ are repeated using the periodic boundary conditions to populate the volume corresponding to the CMASS sample in the sky. We ray-trace the simulated volume following {\sc HEALPix} pixels at $N_{\rm side} = 1024$. The simulated $y$-map is constructed by summing up all the haloes within $0.43 < z < 0.75$ that intersect with each {\sc HEALPix} pixel. The map is then convolved with a Gaussian filter of FWHM 10 arcmin to mimic the \textit{Planck} $y$-map. We note that the effect of gas in small haloes is almost certainly overestimated in this method, as we have extrapolated the $Y-M$ relationship down to a low mass of $10^{10}\,h^{-1}M_{\odot}$.
We know that in reality the gas fraction drops below the universal baryon fraction for low mass haloes, possibly from $10^{13}\,h^{-1}M_{\odot}$ \citep{Lim2017}, and hence the contribution from low mass haloes may have been significantly overestimated given that the number of them is much larger. Crucially, our simulated $y$-map does not include any contribution from a diffuse intergalactic medium, so that any SZ signal detected between two mock CMASS galaxies derives solely from haloes that lie between the pair and along the line of sight.

To construct the corresponding galaxy pair catalogues, we populate the simulated halo catalogues in the CMASS volume with galaxies using an HOD recipe \citep{Peacock2000,Scoccimarro2001,Benson2000,Berlind2002,Kravtsov2004}. The mean occupancy of central and satellite galaxies for each halo are described by a model with five free parameters \citep{Zheng2007}, for which we adopt the values from \citet{Manera2013}. These are similar to those of \citet{White2011} and calibrated to reproduce the clustering of the SDSS DR9 CMASS sample. Despite the slight differences of the cosmological parameters between the Millennium simulation and those used by the above papers, the mean number densities of galaxies are found to be similar, at approximately $3\times10^{-4}(h^{-1}{\rm Mpc})^{-3}$. Satellite galaxies are distributed following an NFW profile \citep{NFW} with the halo concentration parameter from \citet{Neto2007}. The distribution of satellites within the halo is unimportant for this study since (a) the satellite fraction is only $\sim\,$ 10\%; (b) the scale of our concern is well beyond the 1-halo term; and (c) the convolution of the map with the 10-arcmin Gaussian beam erases information on these scales. With the simulated $y$-map and its corresponding galaxy catalogues, we repeat our stacking analysis for all galaxy pairs in the CMASS footprint. The resulting stacked maps and residuals are shown in Fig.~\ref{fig:Millennium_2D_plots}. After subtraction of the isotropic component from the two haloes, we find a small residual filament between the galaxy pairs at the level of $\sim\,$2\% of the peak values of the two haloes. This is consistent with our estimate based on the CMASS galaxy sample shown in Fig.~\ref{fig:num_density}. We therefore estimate that correlated galaxy haloes contribute approximately 20\% of the detected filament signal.

\subsection{Dust contamination}

The \textit{Planck} $y$-map may have been contaminated by the CIB, as speculated by the \textit{Planck} team \citep{planck2016ymaps}. This is a result of the non-zero CIB emission within the SZ-sensitive wavelength range. Whilst the CIB emission from dusty galaxies comes mainly from higher redshifts than the CMASS sample, there is some overlap in the redshift distributions \citep{planck2016ymaps}.
The fraction of the CIB leakage coming from the \textit{Planck} 857~GHz map can be quantified as $\alpha_{\rm CIB}$.
This quantity has been estimated by \citet{Vikram2017} for the same $y$-map used for our study, and by \citet{Hill2014} for a $y$-map constructed independently from \textit{Planck} data, to be approximately $10^{-7}$ and $10^{-6}$ for their respective cross-correlation analysis of SZ$\times$SDSS groups at $z < 0.2$ and SZ$\times$CMB lensing at $z \simeq 2$. With the difference between the redshift distribution of the CMASS galaxies and these studies, we conclude that the fraction of CIB contamination is therefore between $10^{-7} < \alpha_{\rm CIB} <10^{-6}$. When repeating our stacking analysis using the \textit{Planck} 857~GHz map, we find a residual filament signal with an amplitude of the order of $10^{-5}\,$K in terms of the thermal temperature of the CMB \citep{Planck2014IX}. When multiplied by the leakage parameter $\alpha_{\rm CIB}$, the CIB contamination is estimated to be 2-3 orders of magnitudes below the detected SZ filament. 

The above estimates assume that the leakage coefficient $\alpha_{\rm CIB}$ applies to all scales, regardless of the possible difference between the CIB and SZ $y$ power spectra; but an independent study from the Planck collaboration \citep{PlanckCIB}, has shown that the estimated CIB leakage power spectrum is steeper than the power spectrum of $y$. However, the relative amplitude of these two power spectra does not change by more than than one order of magnitude at the scales of our interests ($100<\ell<2000$). Therefore, the order-of-magnitude estimate from Vikram et al. remains valid. Perhaps more importantly, the cross-spectrum between the CIB leakage and $y$ was shown to be one order of magnitude below their auto-spectra at all scales \citep{PlanckCIB}. This suggests that only 10\% of the CIB leakage is correlated with the SZ signal, therefore supporting our conclusion that the CIB leakage to the excess SZ signal is sub-dominant.

Despite the remaining uncertainties in the CIB-SZ leakage, it is possible to make a direct argument that such leakage cannot be the source of our apparent filament signal. This argument exploits the logic of section~\ref{sec:gasinhaloes}, where we considered whether the filament signal could arise from the blended SZ effect of many haloes lying along the filament.
We concluded that this could not be the case, based on the small stacked galaxy density in the filament region, relative to the stacked halo signal around the CMASS galaxies that mark the filament. The filament:halo ratio is larger in the SZ stack, implying that some diffuse source is required for the SZ signal, in addition to the SZ signal from virialised gas in haloes. If this argument is accepted, it is readily seen that CIB leakage must be irrelevant: the CIB is associated with star formation and will only contaminate the apparent SZ signal from haloes. The fraction of the halo SZ signal that is induced by the CIB will have no effect on the contrast between filament centre and ends and thus cannot produce our apparent excess diffuse SZ signal.

\subsection{Estimating the baryon content in filaments}\label{sec:gas_props}

\begin{figure*}[p]
\centering
\includegraphics[width=0.95\linewidth]{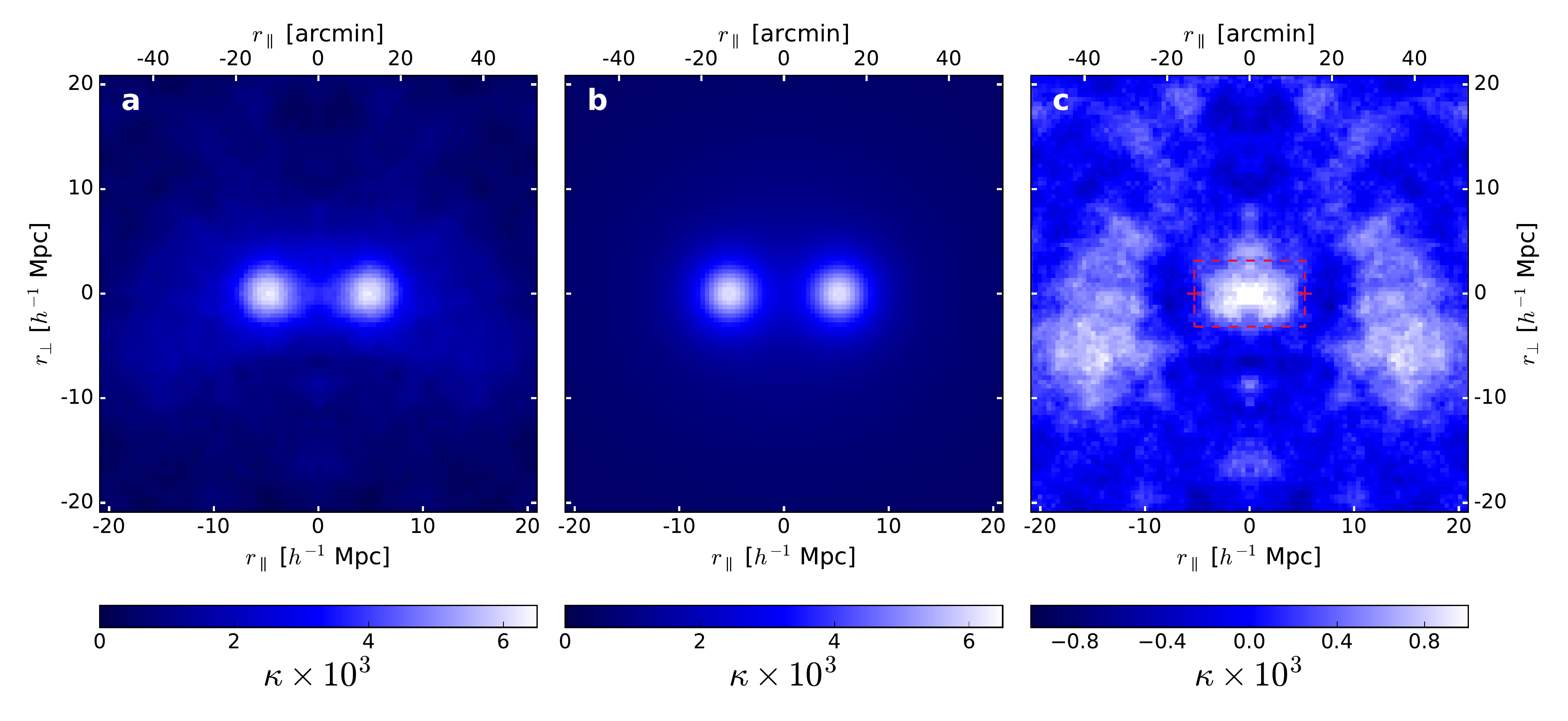}
\caption{Panels represent the same as in Fig.~\ref{fig:2D_plots}, except that the projected total matter density parameter $\kappa$ is now presented. Similar to the results from the SZ $y$-map, after subtracting the isotropic component of the two haloes (b) from the observed signal in (a), we find residual lensing signal between the galaxy pairs shown in (c).} 
\label{fig:lensing_2D_plots}
\end{figure*}

\begin{figure*}[p]
\centering
\includegraphics[width=0.9\linewidth]{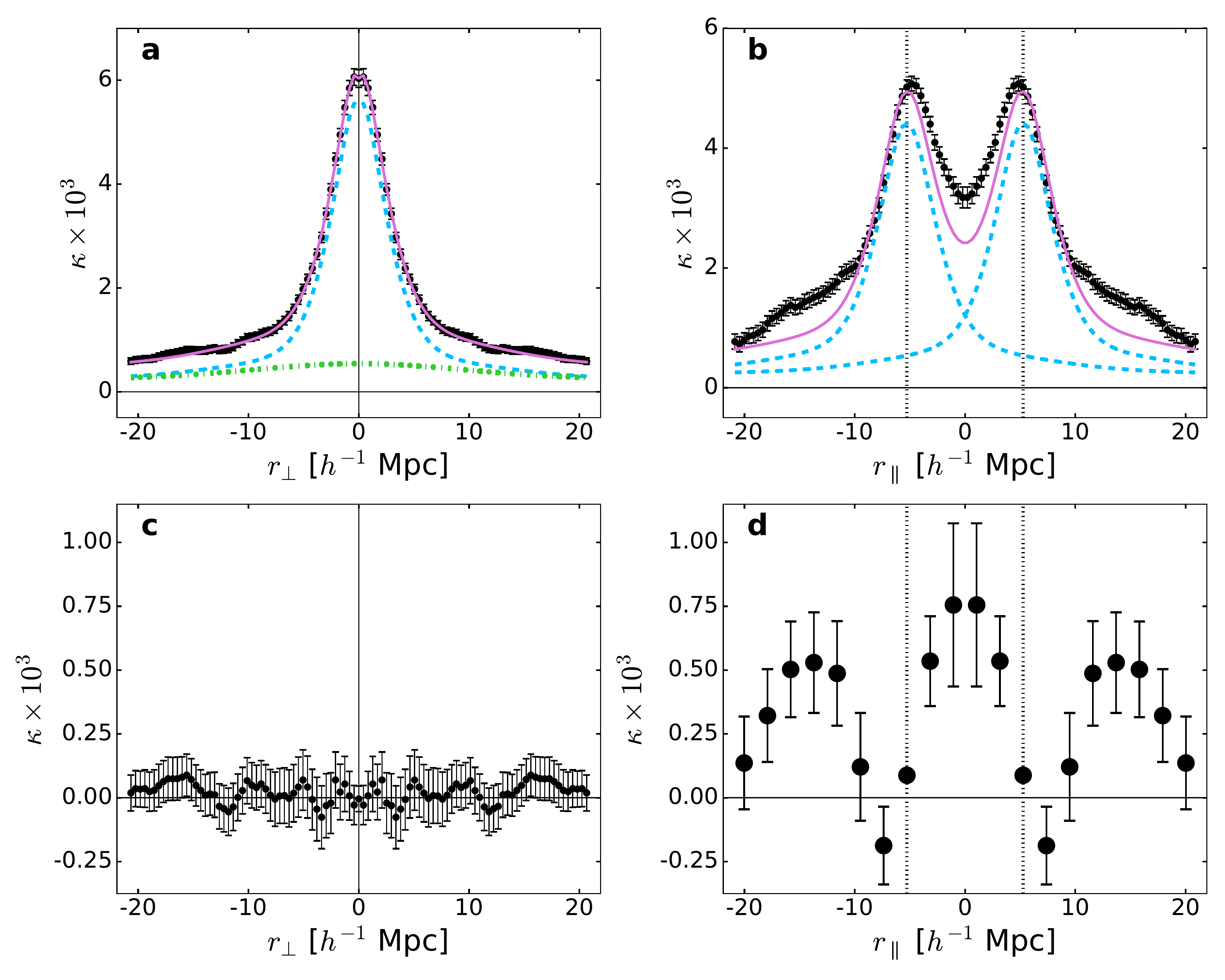}
\caption{One-dimensional profiles of the lensing signal from stacked galaxy pairs. The meaning of each panel is the same as in Fig.~\ref{fig:1D_filament} except that the projected total matter density parameter $\kappa$ is now presented. Error bars in panels a and b represent the errors on the mean, whereas panels c and d show the errors obtained from the jackknife analysis.}
\label{fig:lensing_1D_filament}
\end{figure*}

The SZ effect constrains only the product of gas temperature and gas density (Eq.~\ref{eq:tsz}). From our analysis we measure the mean amplitude of the Compton $y$-parameter to be $\bar y = 0.6\pm 0.2 \times10^{-8}$. To break the degeneracy between the gas density and temperature in this measurement, we require an independent estimate for one of the quantities.

Previous studies of warm-hot gas filaments using hydrodynamical simulations \citep[e.g.][]{hydrosim1999,dave2001,Martizzi2018} have suggested that gas in filaments should have a lower density than gas within virialised objects, with a typical median gas density of order $10 \bar{n}_{\rm e}$. This gas has been shock heated to temperatures ranging between $10^5$ to $10^7 \,$K, following a distribution that peaks at approximately $10^6 \,$K. Assuming this peak temperature for our measured filament, we can estimate its gas density as follows.

We model the filament as a cylinder in which the gas distribution follows the total matter distribution and they both follow a two-dimensional Gaussian profile. Unlike the line-of-sight direction, along the vertical direction ($r_\perp$) the SZ signal and hence the electron number density $n_\mathrm{e}$ will also be convolved with the \textit{Planck} beam of FWHM 10 arcmin. We adopt a FWHM for the intrinsic filament profile of $ 1.5\mpcoh$ \citep{colberg2005}. We note that our result for the total baryon content of the filament is almost independent of this choice. Since the large \textit{Planck} beam is dominant, the chosen intrinsic width makes a difference of $\sim\,$1 percentage point to the baryon content estimate. From our analysis (for a complete derivation, see Appendix~\ref{sec:derivation}) we found the mean amplitude of the filament
between the two pairs to be $y\simeq 0.6\times10^{-8}$. Taking $T_{\rm e}=10^6$K and using the cylindrical Gaussian model, we determine the mean density in the filament to be $n_{\rm e}(z)\simeq6\times \bar{n_{\rm e}}(z)$, where $n_{\rm e}(z)$ is the mean universal electron density at the medium redshift of the CMASS galaxy sample $z=0.55$. Assuming the universe to be fully ionised and accounting for the full volume occupied by the
CMASS galaxies of $\sim\,$4 ($h^{-1}$Gpc)$^3$, this estimated filament density amounts to approximately
30\% of the mean baryon density of the Universe: 0.3$\Omega_b$.

This estimate is based on the mean gas temperature from hydrodynamical simulations, which comes with a large uncertainty, mainly due to limitations on our understanding of galaxy formation and the baryon cycle in the Universe. Moreover, our selection of gas filaments from observations may not be fully representative of the WHIM studied in these simulations. More elaborate comparisons between simulations and observations would need to be conducted to verify the above result. Alternatively however, we can estimate the internal mass density of the filament using gravitational lensing. If we assume that dark matter and baryons have not separated on these scales, we can obtain an estimate of the baryon density, and hence the gas temperature.

We repeat our stacking analysis, replacing the SZ $y$-map with the modified lensing $\kappa$-map from \textit{Planck} (section~\ref{sec:lensingmap}). 
Following the same analysis as for the SZ results, the results of which are shown in Figs~\ref{fig:lensing_2D_plots} \& \ref{fig:lensing_1D_filament}, we measure the mean amplitude of the projected matter density in the filament between the galaxy pairs to be $\bar \kappa = (0.58\pm 0.31) \times 10^{-3}$, a $1.9\sigma$ measurement. Including the excess signal outside the galaxy pairs, we find a $3.1\sigma$ detection. This is undeniably a marginal signal, but the detection of the imprint of filaments on CMB lensing has previously also been reported in \citet{He2017} using a different method.

Using the same cylindrical model as for the SZ measurement (see Appendix~\ref{sec:derivation}), we estimate the central total matter density in the filament to be {$\rho(z)= (5.5 \pm 2.9) \times \bar{\rho}(z)$, where $\bar{\rho} (z)$ is the mean universal total matter density at the median redshift of the CMASS galaxy sample $z=0.55$.}
Assuming that the baryon fraction of the filament follows the mean value of the Universe (i.e. the gas density $n_{\rm 0}(z) =(5.5 \pm 2.9)\times \bar{n}_{\rm e}(z)$), the gas temperature is estimated to be {$T_{\rm e}=(2.7 \pm 1.7)\times10^6 \,$K}, where we assume that the errors between the lensing and SZ measurements are independent. Given the relatively large uncertainties from both simulations and observations, we find our constraints for the gas density and temperature of the filament to be broadly consistent with expectations from the literature.

Combining this gas density and temperature, the gas in our sample of filaments amounts to $(0.11\pm 0.07) \Omega_{\rm b}$. We note that the dominant source of contamination of this signal arises from gas in haloes within the filament. This was found to contribute less than 20\% of the total signal, a level of contamination that is within the overall error on our measurement.

The above estimates use only the signal measured in the region interior to the galaxy pairs (i.e. the boxed region indicated in Figs.~\ref{fig:2D_plots} \& \ref{fig:lensing_2D_plots}), but in section~\ref{sec:covariance} we found the filament signal to be of higher significance when when we extend the filament region beyond the galaxy pairs. The mean convergence of this wider area is $\bar \kappa = (0.33\pm 0.11) \times 10^{-3}$, with a mean amplitude of the Compton $y$-parameter of $\bar y = 2.4\pm 0.6 \times10^{-9}$. Performing the same calculations as before to estimate the baryon content for filaments spanning the full width of the stacked map ($42\mpcoh$), we find the central total matter density in the filament to be $\rho(z)= (3.5 \pm 1.1) \times \bar{\rho}(z)$ and the temperature to be $T_{\rm e}=(1.7 \pm 0.7)\times10^6 \,$K, corresponding to a baryon density of $(0.28\pm 0.12) \Omega_{\rm b}$. 
The maximum spatial extent of the filaments is not known a priori, but the horizontal residuals for the SZ and lensing measurement do both converge to zero towards the edge of the region (lower panels of Figs. \ref{fig:1D_filament} \& \ref{fig:lensing_1D_filament}), indicating a possible natural length-scale of the filament. 
Nevertheless, because of the slight uncertainty over the choice of filament length, we generally regard the estimates obtained using only the region interior to the galaxy pairs as our main result.

\begin{figure}

\centering
\includegraphics[width=\linewidth]{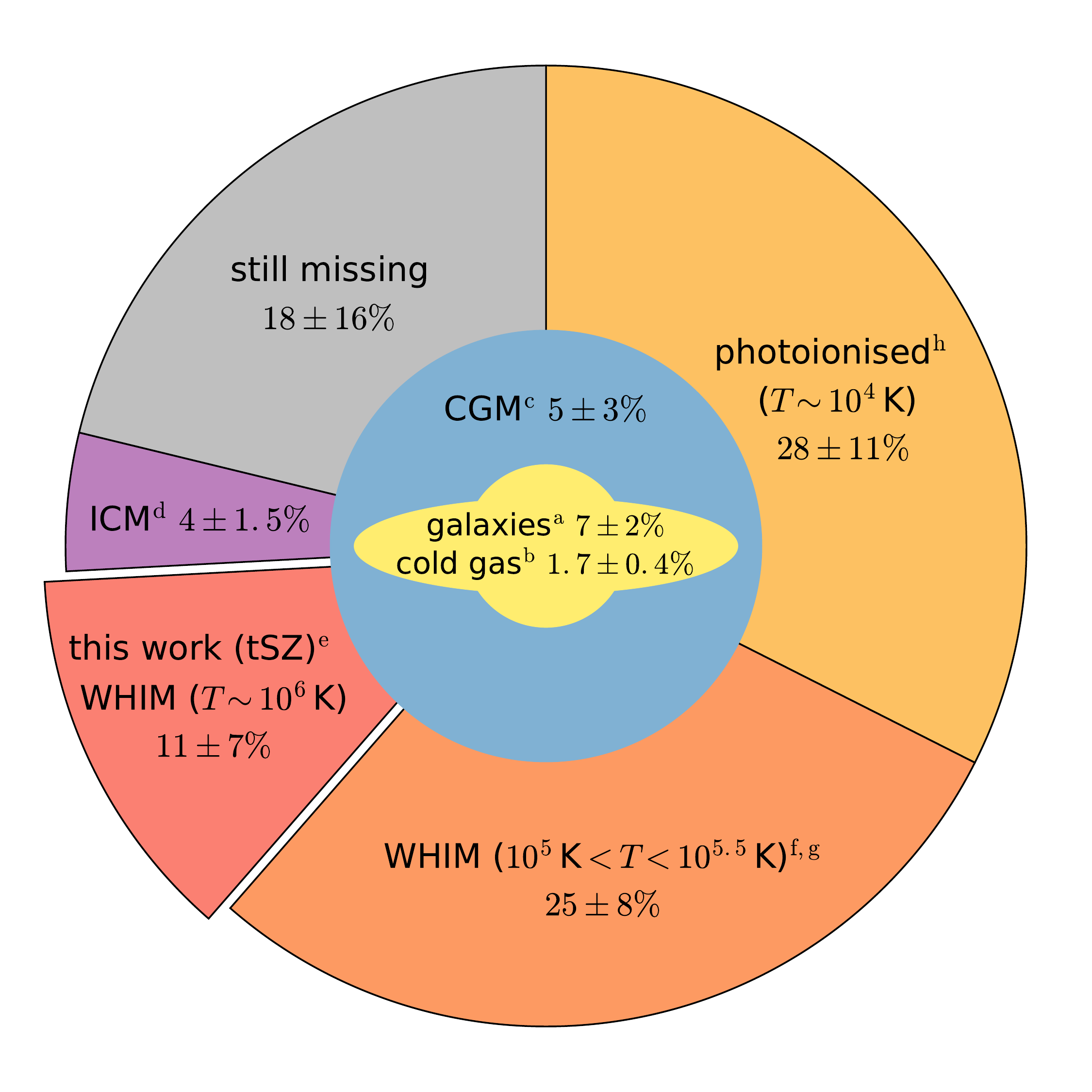}
\caption{Summary of the baryon budget of the late-time Universe.  Table values are adjusted from  \citet{Fukugita2004}, \citet{shull2012} and \citet{Nicastro2017}. 
(a) stellar component of galaxies; (b) cold gas in galaxies estimated from observation of HI and $\rm {H_{\rm 2}}$; (c) circum-galactic medium observed through $\rm{Ly \alpha}$ O VI absorptions\citep{Prochaska2011};
(d) intracluster medium observed via X-ray emission. The recent detection of X-ray filaments near a massive nearby cluster \citep{xray:filaments}, and SZ signal between interacting clusters \citep{PlanckFilament} measures hot gas in a similar phase as in this category. 
(e) warm-hot intergalactic medium (WHIM) detected via the thermal SZ effect; the value is estimated at $z\simeq0.5$, which is consistent with the estimate by \citet{Nicastro2018} from the observation of two OVII absorbers; (f) and (g) WHIM observed via Damped $\rm{Ly \alpha}$ and O VI absorptions; \citet{shull2012} noted that there is some overlap between DLA and O VI. According to \citet{Nicastro2017}, they trace the same gas. This is the biggest uncertainty for the amount of baryons which is still missing. (h) gas observed via photonionised $\rm{Ly \alpha}$ absorbers. Note that the recent detection of the kinetic SZ effect \citep{kSZ2015HM, Hill2016, Schaan2016, DeBernardis2017} constrains the baryon content in and around collapsed objects, and therefore falls into the category of (c) and (d).}
\label{fig:PieChart}
\end{figure}

\section{Discussion}\label{sec:discussion}

The quoted baryon fraction refers to gas found in filaments derived from our specific selection in terms of filament length and galaxy pair population, so it is certainly incomplete. We expect that more gas in filaments can be detected using deeper galaxy surveys, in which smaller filaments will be found. An increased sample size can, together with future lensing and X-ray surveys, also provide better constraints on the density and temperature, and hence on the inferred baryon fraction. For the present, in Fig.~\ref{fig:PieChart} we summarise the current census of baryons in the low-redshift Universe, combining our work with existing literature \citep{shull2012,Nicastro2017}.

Our study of gas filaments using the thermal SZ effect is the latest installment in a long campaign to search for the missing baryons in the intergalactic medium. Direct observations have been made through Ly$\alpha$, OVII and OVI absorption lines in quasar spectra \citep{penton2004, Nicastro2008, Tejos2016,Nicastro2018}. \citet{Nicastro2018} determined an upper and lower limit on the warm-hot baryon fraction from the measurement of two OVII absorbers. Our result is in agreement with their estimate, however the uncertainties in both studies are very large, such that no strong conclusions on the baryon fraction can be drawn.

Several previous works have focused on the X-ray emission and SZ signal from individual filaments. \citet{kull1999} reported the presence of an elongated gas structure within the Shapley Supercluster in \textit{ROSAT} data. \citet{xray:filaments} detected multiple filamentary structures around the Abell 2744 cluster using XMM-Newton observations. The gas in these large-scale ($\sim\,$$1-2\mpcoh$) filaments was found to be at temperatures $T\simeq 10^7\,$K and densities $\rho_{\rm b} \simeq 200\bar{\rho_{\rm b}}$, and therefore represent the high-temperature and high-density end of the WHIM. 

The \citet{PlanckFilament} combined tSZ data and \textit{ROSAT} X-ray data to study pairs of merging clusters. They detect a filamentary structure of projected length $3\mpcoh$ connecting the cluster pair A399-A401, finding the Compton parameter to be $y\sim10^5$, and estimate the gas temperature to be $k_{\rm B}T=7.08\pm0.85\,\rm keV$ ($T\sim8\times10^7\,$K) and the electron density to be $n_{\rm e}=3.7\times10^{-4}\,\rm cm^{-3}$ (over-density $\delta\sim200$).
\citet{Bonjean2017} performed a similar analysis, however assuming a different X-ray temperature measurement, and found a similar Compton $y$-parameter and electron density of $n_{\rm e}=(4.3\pm0.7)\times10^{-4}\,\rm cm^{-3}$.

Our work clearly probes a different regime than these previous studies, detecting large-scale structures of baryons that span $\sim\,$$10\mpcoh$ at a significantly lower density ($\sim\,$$5\rho_b$) and lower temperature ($\sim\,$$10^6 \,$K). Moreover, in comparison with previous absorption line studies, our method probes an unprecedented cosmological volume and provides a visualisation of the WHIM.

Similar conclusions to this work have been reached independently by \citet{tanimura2017} (hereafter referred to as T19).
Our study differs from T19 mainly in the galaxy pair catalogues used: T19 used the SDSS-DR12 LRG galaxy catalogue and found 262,864 pairs of galaxies at redshifts $z<0.4$. We used the DR12 CMASS galaxy catalogue and found 1 million pairs with similar selection criteria. Our sample is 5 times larger and covers a higher redshift range ($0.43<z<0.75$). These two catalogues are therefore independent and complementary in their redshift ranges.  Despite the differences, we achieved similar results in terms of the amplitudes and statistical significances of the filament signal. In terms of the Compton $y$-parameter, T19 found $y \simeq 1\times10^{-8}$ at the $5.3\sigma$ level, whereas we find $y \simeq 0.6\times10^{-8}$ at the {$2.9\sigma$} level and $y\simeq 0.24\times10^{-8}$ at the $3.8\sigma$ level when also including the signal beyond the galaxy pairs. The fact that two independent studies using two different catalogues achieve similar conclusions provides strong evidence for the detection of gas filaments.

\section{Conclusions}\label{sec:conclusion}

Cosmological simulations predict that the majority of missing baryons in the Universe form a diffuse warm-hot gas in filaments in the cosmic web. We searched for gas filaments by stacking the thermal Sunyaev-Zel'dovich $y$-map from \textit{Planck} for 1,002,334 pairs of CMASS galaxies with a mean comoving separation of $10.5\mpcoh$. We found a significant SZ signal for the stacked galaxy pairs, and modelled the galaxy pair haloes by assuming their pressure profile to be isotropic. After subtraction of the galaxy pair contribution to the signal, we find residual filamentary signal at the $2.9\sigma$ confidence level with a mean Compton parameter $\bar{y}=(0.6\pm0.2)\times 10^{-8}$. 

We have identified and addressed potential sources of contamination to this residual signal, including contributions from uncorrelated sources and dust in the Milky Way, bound gas in haloes correlated with the galaxy pairs, and systematic leakage from the CIB into the SZ $y$-map. We conclude that the only significant source of contamination is due to bound gas in correlated structures, which may contribute up to $20\%$ of the measured residual signal. 

To estimate the filament gas properties we measured their gravitational lensing signal by stacking the \textit{Planck} $\kappa$-map for the same sample of galaxy pairs, finding a mean convergence $\kappa=(0.6\pm0.3)\times10^{-3}$, a marginal detection of $1.9\sigma$. Assuming a cylindrical filament model in which both the baryons and dark matter follow a Gaussian profile, we obtain a filament gas density $\rho_{\rm b}=(5.5\pm2.9)\times\bar{\rho_{\rm b}}$ and temperature  $T=(2.7\pm 1.7) \times 10^6\,$K, which (albeit with large uncertainty) is consistent with predictions from hydrodynamical simulations. Considering the total volume spanned by the CMASS galaxies, this measurement can account for $11\pm7\%$ of the total baryon content of the Universe. 

Additionally, we have found evidence that the filaments extend beyond the galaxy pairs. Including the contribution from this extended region increases the significance of the filament signal to $3.8\sigma$ and $3.1\sigma$ for the SZ and lensing measurement respectively, and overall this extended filament region may account for $28\pm 12\%$ of the total baryon content.

Our method complements other probes for the intergalactic medium such as X-ray and quasar absorption; it provides evidence for the presence of the WHIM in filaments and opens up a new window to search for missing baryons in the cosmic web. 

\begin{acknowledgements}
We thank Martin White, Vinu Vikram, and Adam Lidz for useful discussions. We thank Carlos Hern\'andez-Monteagudo for useful comments on an early version of our paper. AdG was supported by the Edinburgh School of Physics and Astronomy Career Development Summer Scholarship and the RSE Cormack Vacation Research Scholarship. YC, JAP, and CH were supported by the European Research Council under grant numbers 670193 (YC; JAP) and 647112 (CH). 
We thank the Planck collaboration for making the full-sky $y$-map and $\kappa$-map publicly available and the SDSS collaboration for making public the CMASS galaxy catalogue. 
Funding for SDSS-III has been provided by the Alfred P. Sloan Foundation, the Participating Institutions, the National Science Foundation, and the U.S. Department of Energy Office of Science. The SDSS-III web site is {http://www.sdss3.org/}.

SDSS-III is managed by the Astrophysical Research Consortium for the Participating Institutions of the SDSS-III Collaboration including the University of Arizona, the Brazilian Participation Group, Brookhaven National Laboratory, Carnegie Mellon University, University of Florida, the French Participation Group, the German Participation Group, Harvard University, the Instituto de Astrofisica de Canarias, the Michigan State/Notre Dame/JINA Participation Group, Johns Hopkins University, Lawrence Berkeley National Laboratory, Max Planck Institute for Astrophysics, Max Planck Institute for Extraterrestrial Physics, New Mexico State University, New York University, Ohio State University, Pennsylvania State University, University of Portsmouth, Princeton University, the Spanish Participation Group, University of Tokyo, University of Utah, Vanderbilt University, University of Virginia, University of Washington, and Yale University.

\end{acknowledgements}

\bibliography{bibfile}

\begin{thebibliography}{62}
\expandafter\ifx\csname natexlab\endcsname\relax\def\natexlab#1{#1}\fi

\bibitem[{{Alam} {et~al.}(2015){Alam}, {Albareti}, {Allende Prieto}, {Anders},
  {Anderson}, {Anderton}, {Andrews}, {Armengaud}, {Aubourg}, {Bailey}, \&
  et~al.}]{sdss:dr12}
{Alam}, S., {Albareti}, F.~D., {Allende Prieto}, C., {et~al.} 2015, \apjs, 219,
  12

\bibitem[{{Benson} {et~al.}(2000){Benson}, {Cole}, {Frenk}, {Baugh}, \&
  {Lacey}}]{Benson2000}
{Benson}, A.~J., {Cole}, S., {Frenk}, C.~S., {Baugh}, C.~M., \& {Lacey}, C.~G.
  2000, \mnras, 311, 793

\bibitem[{{Berlind} \& {Weinberg}(2002)}]{Berlind2002}
{Berlind}, A.~A. \& {Weinberg}, D.~H. 2002, \apj, 575, 587

\bibitem[{{Bonamente} {et~al.}(2008){Bonamente}, {Joy}, {LaRoque}, {Carlstrom},
  {Nagai}, \& {Marrone}}]{Bonamente2008}
{Bonamente}, M., {Joy}, M., {LaRoque}, S.~J., {et~al.} 2008, \apj, 675, 106

\bibitem[{{Bonjean} {et~al.}(2017){Bonjean}, {Aghanim}, {Salom{\'e}},
  {Douspis}, \& {Beelen}}]{Bonjean2017}
{Bonjean}, V., {Aghanim}, N., {Salom{\'e}}, P., {Douspis}, M., \& {Beelen}, A.
  2017, ArXiv e-prints [\eprint[arXiv]{1710.08699}]

\bibitem[{{Cen} \& {Ostriker}(1999)}]{hydrosim1999}
{Cen}, R. \& {Ostriker}, J.~P. 1999, \apj, 514, 1

\bibitem[{{Clampitt} {et~al.}(2016){Clampitt}, {Miyatake}, {Jain}, \&
  {Takada}}]{Clampitt2016}
{Clampitt}, J., {Miyatake}, H., {Jain}, B., \& {Takada}, M. 2016, \mnras, 457,
  2391

\bibitem[{{Colberg} {et~al.}(2005){Colberg}, {Krughoff}, \&
  {Connolly}}]{colberg2005}
{Colberg}, J.~M., {Krughoff}, K.~S., \& {Connolly}, A.~J. 2005, \mnras, 359,
  272

\bibitem[{{Cyburt} {et~al.}(2016){Cyburt}, {Fields}, {Olive}, \&
  {Yeh}}]{Cyburt2016}
{Cyburt}, R.~H., {Fields}, B.~D., {Olive}, K.~A., \& {Yeh}, T.-H. 2016, Reviews
  of Modern Physics, 88, 015004

\bibitem[{{da Silva} {et~al.}(2004){da Silva}, {Kay}, {Liddle}, \&
  {Thomas}}]{daSilva2004}
{da Silva}, A.~C., {Kay}, S.~T., {Liddle}, A.~R., \& {Thomas}, P.~A. 2004,
  \mnras, 348, 1401

\bibitem[{{Dav{\'e}} {et~al.}(2001){Dav{\'e}}, {Cen}, {Ostriker}, {Bryan},
  {Hernquist}, {Katz}, {Weinberg}, {Norman}, \& {O'Shea}}]{dave2001}
{Dav{\'e}}, R., {Cen}, R., {Ostriker}, J.~P., {et~al.} 2001, \apj, 552, 473

\bibitem[{{Dawson} {et~al.}(2013){Dawson}, {Schlegel}, {Ahn}, {Anderson},
  {Aubourg}, {Bailey}, {Barkhouser}, {Bautista}, {Beifiori}, {Berlind},
  {Bhardwaj}, {Bizyaev}, {Blake}, {Blanton}, {Blomqvist}, {Bolton}, {Borde},
  {Bovy}, {Brandt}, {Brewington}, {Brinkmann}, {Brown}, {Brownstein}, {Bundy},
  {Busca}, {Carithers}, {Carnero}, {Carr}, {Chen}, {Comparat}, {Connolly},
  {Cope}, {Croft}, {Cuesta}, {da Costa}, {Davenport}, {Delubac}, {de Putter},
  {Dhital}, {Ealet}, {Ebelke}, {Eisenstein}, {Escoffier}, {Fan}, {Filiz Ak},
  {Finley}, {Font-Ribera}, {G{\'e}nova-Santos}, {Gunn}, {Guo}, {Haggard},
  {Hall}, {Hamilton}, {Harris}, {Harris}, {Ho}, {Hogg}, {Holder}, {Honscheid},
  {Huehnerhoff}, {Jordan}, {Jordan}, {Kauffmann}, {Kazin}, {Kirkby}, {Klaene},
  {Kneib}, {Le Goff}, {Lee}, {Long}, {Loomis}, {Lundgren}, {Lupton}, {Maia},
  {Makler}, {Malanushenko}, {Malanushenko}, {Mandelbaum}, {Manera}, {Maraston},
  {Margala}, {Masters}, {McBride}, {McDonald}, {McGreer}, {McMahon}, {Mena},
  {Miralda-Escud{\'e}}, {Montero-Dorta}, {Montesano}, {Muna}, {Myers},
  {Naugle}, {Nichol}, {Noterdaeme}, {Nuza}, {Olmstead}, {Oravetz}, {Oravetz},
  {Owen}, {Padmanabhan}, {Palanque-Delabrouille}, {Pan}, {Parejko},
  {P{\^a}ris}, {Percival}, {P{\'e}rez-Fournon}, {P{\'e}rez-R{\`a}fols},
  {Petitjean}, {Pfaffenberger}, {Pforr}, {Pieri}, {Prada}, {Price-Whelan},
  {Raddick}, {Rebolo}, {Rich}, {Richards}, {Rockosi}, {Roe}, {Ross}, {Ross},
  {Rossi}, {Rubi{\~n}o-Martin}, {Samushia}, {S{\'a}nchez}, {Sayres}, {Schmidt},
  {Schneider}, {Sc{\'o}ccola}, {Seo}, {Shelden}, {Sheldon}, {Shen}, {Shu},
  {Slosar}, {Smee}, {Snedden}, {Stauffer}, {Steele}, {Strauss}, {Streblyanska},
  {Suzuki}, {Swanson}, {Tal}, {Tanaka}, {Thomas}, {Tinker}, {Tojeiro},
  {Tremonti}, {Vargas Maga{\~n}a}, {Verde}, {Viel}, {Wake}, {Watson}, {Weaver},
  {Weinberg}, {Weiner}, {West}, {White}, {Wood-Vasey}, {Yeche}, {Zehavi},
  {Zhao}, \& {Zheng}}]{boss:dr12}
{Dawson}, K.~S., {Schlegel}, D.~J., {Ahn}, C.~P., {et~al.} 2013, \aj, 145, 10

\bibitem[{{De Bernardis} {et~al.}(2017){De Bernardis}, {Aiola}, {Vavagiakis},
  {Battaglia}, {Niemack}, {Beall}, {Becker}, {Bond}, {Calabrese}, {Cho},
  {Coughlin}, {Datta}, {Devlin}, {Dunkley}, {Dunner}, {Ferraro}, {Fox},
  {Gallardo}, {Halpern}, {Hand}, {Hasselfield}, {Henderson}, {Hill}, {Hilton},
  {Hilton}, {Hincks}, {Hlozek}, {Hubmayr}, {Huffenberger}, {Hughes}, {Irwin},
  {Koopman}, {Kosowsky}, {Li}, {Louis}, {Lungu}, {Madhavacheril}, {Maurin},
  {McMahon}, {Moodley}, {Naess}, {Nati}, {Newburgh}, {Nibarger}, {Page},
  {Partridge}, {Schaan}, {Schmitt}, {Sehgal}, {Sievers}, {Simon}, {Spergel},
  {Staggs}, {Stevens}, {Thornton}, {van Engelen}, {Van Lanen}, \&
  {Wollack}}]{DeBernardis2017}
{De Bernardis}, F., {Aiola}, S., {Vavagiakis}, E.~M., {et~al.} 2017, \jcap, 3,
  008

\bibitem[{{Eckert} {et~al.}(2015){Eckert}, {Jauzac}, {Shan}, \&
  et~al.}]{xray:filaments}
{Eckert}, D., {Jauzac}, M., {Shan}, H., \& et~al. 2015, \nat, 528, 105

\bibitem[{{Epps} \& {Hudson}(2017)}]{hudson2017}
{Epps}, S.~D. \& {Hudson}, M.~J. 2017, \mnras, 468, 2605

\bibitem[{{Fukugita} \& {Peebles}(2004)}]{Fukugita2004}
{Fukugita}, M. \& {Peebles}, P.~J.~E. 2004, \apj, 616, 643

\bibitem[{{G{\'o}rski} {et~al.}(2005){G{\'o}rski}, {Hivon}, {Banday},
  {Wandelt}, {Hansen}, {Reinecke}, \& {Bartelmann}}]{gorski2005}
{G{\'o}rski}, K.~M., {Hivon}, E., {Banday}, A.~J., {et~al.} 2005, \apj, 622,
  759

\bibitem[{{Greco} {et~al.}(2015){Greco}, {Hill}, {Spergel}, \&
  {Battaglia}}]{Greco2015}
{Greco}, J.~P., {Hill}, J.~C., {Spergel}, D.~N., \& {Battaglia}, N. 2015, \apj,
  808, 151

\bibitem[{{Hartlap} {et~al.}(2007){Hartlap}, {Simon}, \&
  {Schneider}}]{Hartlap2007}
{Hartlap}, J., {Simon}, P., \& {Schneider}, P. 2007, \aap, 464, 399

\bibitem[{{He} {et~al.}(2017){He}, {Alam}, {Ferraro}, {Chen}, \& {Ho}}]{He2017}
{He}, S., {Alam}, S., {Ferraro}, S., {Chen}, Y.-C., \& {Ho}, S. 2017, ArXiv
  e-prints [\eprint[arXiv]{1709.02543}]

\bibitem[{{Hern{\'a}ndez-Monteagudo} {et~al.}(2015){Hern{\'a}ndez-Monteagudo},
  {Ma}, {Kitaura}, {Wang}, {G{\'e}nova-Santos}, {Mac{\'{\i}}as-P{\'e}rez}, \&
  {Herranz}}]{kSZ2015HM}
{Hern{\'a}ndez-Monteagudo}, C., {Ma}, Y.-Z., {Kitaura}, F.~S., {et~al.} 2015,
  Physical Review Letters, 115, 191301

\bibitem[{{Hill} {et~al.}(2016){Hill}, {Ferraro}, {Battaglia}, {Liu}, \&
  {Spergel}}]{Hill2016}
{Hill}, J.~C., {Ferraro}, S., {Battaglia}, N., {Liu}, J., \& {Spergel}, D.~N.
  2016, Physical Review Letters, 117, 051301

\bibitem[{{Hill} \& {Spergel}(2014)}]{Hill2014}
{Hill}, J.~C. \& {Spergel}, D.~N. 2014, \jcap, 2, 030

\bibitem[{{Hurier} {et~al.}(2013){Hurier}, {Mac{\'{\i}}as-P{\'e}rez}, \&
  {Hildebrandt}}]{hurier2013}
{Hurier}, G., {Mac{\'{\i}}as-P{\'e}rez}, J.~F., \& {Hildebrandt}, S. 2013,
  \aap, 558, A118

\bibitem[{{Jenkins} {et~al.}(1998){Jenkins}, {Frenk}, {Pearce}, {Thomas},
  {Colberg}, {White}, {Couchman}, {Peacock}, {Efstathiou}, \&
  {Nelson}}]{Jenkins1998}
{Jenkins}, A., {Frenk}, C.~S., {Pearce}, F.~R., {et~al.} 1998, \apj, 499, 20

\bibitem[{{Kay} {et~al.}(2012){Kay}, {Peel}, {Short}, {Thomas}, {Young},
  {Battye}, {Liddle}, \& {Pearce}}]{Kay2012}
{Kay}, S.~T., {Peel}, M.~W., {Short}, C.~J., {et~al.} 2012, \mnras, 422, 1999

\bibitem[{{Kravtsov} {et~al.}(2004){Kravtsov}, {Berlind}, {Wechsler}, {Klypin},
  {Gottl{\"o}ber}, {Allgood}, \& {Primack}}]{Kravtsov2004}
{Kravtsov}, A.~V., {Berlind}, A.~A., {Wechsler}, R.~H., {et~al.} 2004, \apj,
  609, 35

\bibitem[{{Kull} \& {B{\"o}hringer}(1999)}]{kull1999}
{Kull}, A. \& {B{\"o}hringer}, H. 1999, \aap, 341, 23

\bibitem[{{Lim} {et~al.}(2017){Lim}, {Mo}, {Li}, {Liu}, {Ma}, {Wang}, \&
  {Yang}}]{Lim2017}
{Lim}, S., {Mo}, H., {Li}, R., {et~al.} 2017, ArXiv e-prints
  [\eprint[arXiv]{1710.06856}]

\bibitem[{{Manera} {et~al.}(2013){Manera}, {Scoccimarro}, {Percival},
  {Samushia}, {McBride}, {Ross}, {Sheth}, {White}, {Reid}, {S{\'a}nchez}, {de
  Putter}, {Xu}, {Berlind}, {Brinkmann}, {Maraston}, {Nichol}, {Montesano},
  {Padmanabhan}, {Skibba}, {Tojeiro}, \& {Weaver}}]{Manera2013}
{Manera}, M., {Scoccimarro}, R., {Percival}, W.~J., {et~al.} 2013, \mnras, 428,
  1036

\bibitem[{{Maraston} {et~al.}(2013){Maraston}, {Pforr}, {Henriques}, {Thomas},
  {Wake}, {Brownstein}, {Capozzi}, {Tinker}, {Bundy}, {Skibba}, {Beifiori},
  {Nichol}, {Edmondson}, {Schneider}, {Chen}, {Masters}, {Steele}, {Bolton},
  {York}, {Weaver}, {Higgs}, {Bizyaev}, {Brewington}, {Malanushenko},
  {Malanushenko}, {Snedden}, {Oravetz}, {Pan}, {Shelden}, \&
  {Simmons}}]{Maraston2013}
{Maraston}, C., {Pforr}, J., {Henriques}, B.~M., {et~al.} 2013, \mnras, 435,
  2764

\bibitem[{{Martizzi} {et~al.}(2018){Martizzi}, {Vogelsberger}, {Artale},
  {Haider}, {Torrey}, {Marinacci}, {Nelson}, {Pillepich}, {Weinberger},
  {Hernquist}, {Naiman}, \& {Springel}}]{Martizzi2018}
{Martizzi}, D., {Vogelsberger}, M., {Artale}, M.~C., {et~al.} 2018, ArXiv
  e-prints [\eprint[arXiv]{1810.01883}]

\bibitem[{{McGaugh} {et~al.}(2010){McGaugh}, {Schombert}, {de Blok}, \&
  {Zagursky}}]{McGaugh2010}
{McGaugh}, S.~S., {Schombert}, J.~M., {de Blok}, W.~J.~G., \& {Zagursky}, M.~J.
  2010, \apjl, 708, L14

\bibitem[{{Motl} {et~al.}(2005){Motl}, {Hallman}, {Burns}, \&
  {Norman}}]{Motl2005}
{Motl}, P.~M., {Hallman}, E.~J., {Burns}, J.~O., \& {Norman}, M.~L. 2005,
  \apjl, 623, L63

\bibitem[{{Mroczkowski} {et~al.}(2018){Mroczkowski}, {Nagai}, {Basu}, {Chluba},
  {Sayers}, {Adam}, {Churazov}, {Crites}, {Di Mascolo}, {Eckert},
  {Macias-Perez}, {Mayet}, {Perotto}, {Pointecouteau}, {Romero}, {Ruppin},
  {Scannapieco}, \& {ZuHone}}]{Mroczkowski2018}
{Mroczkowski}, T., {Nagai}, D., {Basu}, K., {et~al.} 2018, arXiv e-prints,
  arXiv:1811.02310

\bibitem[{{Navarro} {et~al.}(1996){Navarro}, {Frenk}, \& {White}}]{NFW}
{Navarro}, J.~F., {Frenk}, C.~S., \& {White}, S.~D.~M. 1996, \apj, 462, 563

\bibitem[{{Neto} {et~al.}(2007){Neto}, {Gao}, {Bett}, {Cole}, {Navarro},
  {Frenk}, {White}, {Springel}, \& {Jenkins}}]{Neto2007}
{Neto}, A.~F., {Gao}, L., {Bett}, P., {et~al.} 2007, \mnras, 381, 1450

\bibitem[{{Nicastro} {et~al.}(2018){Nicastro}, {Kaastra}, {Krongold},
  {Borgani}, {Branchini}, {Cen}, {Dadina}, {Danforth}, {Elvis}, {Fiore},
  {Gupta}, {Mathur}, {Mayya}, {Paerels}, {Piro}, {Rosa-Gonzalez}, {Schaye},
  {Shull}, {Torres-Zafra}, {Wijers}, \& {Zappacosta}}]{Nicastro2018}
{Nicastro}, F., {Kaastra}, J., {Krongold}, Y., {et~al.} 2018, \nat, 558, 406

\bibitem[{{Nicastro} {et~al.}(2017){Nicastro}, {Krongold}, {Mathur}, \&
  {Elvis}}]{Nicastro2017}
{Nicastro}, F., {Krongold}, Y., {Mathur}, S., \& {Elvis}, M. 2017,
  Astronomische Nachrichten, 338, 281

\bibitem[{{Nicastro} {et~al.}(2008){Nicastro}, {Mathur}, \&
  {Elvis}}]{Nicastro2008}
{Nicastro}, F., {Mathur}, S., \& {Elvis}, M. 2008, Science, 319, 55

\bibitem[{{Peacock} \& {Smith}(2000)}]{Peacock2000}
{Peacock}, J.~A. \& {Smith}, R.~E. 2000, \mnras, 318, 1144

\bibitem[{{Penton} {et~al.}(2004){Penton}, {Stocke}, \& {Shull}}]{penton2004}
{Penton}, S.~V., {Stocke}, J.~T., \& {Shull}, J.~M. 2004, \apjs, 152, 29

\bibitem[{{Persic} \& {Salucci}(1992)}]{persic1992}
{Persic}, M. \& {Salucci}, P. 1992, \mnras, 258, 14P

\bibitem[{{Planck Collaboration}(2013)}]{PlanckFilament}
{Planck Collaboration}. 2013, \aap, 550, A134

\bibitem[{{Planck Collaboration}(2014)}]{Planck2014IX}
{Planck Collaboration}. 2014, \aap, 571, A9

\bibitem[{{Planck Collaboration}(2016{\natexlab{a}})}]{planck2016ymaps}
{Planck Collaboration}. 2016{\natexlab{a}}, \aap, 594, A22

\bibitem[{{Planck Collaboration}(2016{\natexlab{b}})}]{PlanckCosmology}
{Planck Collaboration}. 2016{\natexlab{b}}, \aap, 594, A13

\bibitem[{{Planck Collaboration}(2016{\natexlab{c}})}]{Plancklensing}
{Planck Collaboration}. 2016{\natexlab{c}}, \aap, 594, A15

\bibitem[{{Planck Collaboration}(2016{\natexlab{d}})}]{PlanckCIB}
{Planck Collaboration}. 2016{\natexlab{d}}, \aap, 594, A23

\bibitem[{{Planck Collaboration}(2016{\natexlab{e}})}]{KSZ_Plannck2016}
{Planck Collaboration}. 2016{\natexlab{e}}, \aap, 586, A140

\bibitem[{{Prochaska} {et~al.}(2011){Prochaska}, {Weiner}, {Chen}, {Mulchaey},
  \& {Cooksey}}]{Prochaska2011}
{Prochaska}, J.~X., {Weiner}, B., {Chen}, H.-W., {Mulchaey}, J., \& {Cooksey},
  K. 2011, \apj, 740, 91

\bibitem[{{Schaan} {et~al.}(2016){Schaan}, {Ferraro}, {Vargas-Maga{\~n}a},
  {Smith}, {Ho}, {Aiola}, {Battaglia}, {Bond}, {De Bernardis}, {Calabrese},
  {Cho}, {Devlin}, {Dunkley}, {Gallardo}, {Hasselfield}, {Henderson}, {Hill},
  {Hincks}, {Hlozek}, {Hubmayr}, {Hughes}, {Irwin}, {Koopman}, {Kosowsky},
  {Li}, {Louis}, {Lungu}, {Madhavacheril}, {Maurin}, {McMahon}, {Moodley},
  {Naess}, {Nati}, {Newburgh}, {Niemack}, {Page}, {Pappas}, {Partridge},
  {Schmitt}, {Sehgal}, {Sherwin}, {Sievers}, {Spergel}, {Staggs}, {van
  Engelen}, {Wollack}, \& {ACTPol Collaboration}}]{Schaan2016}
{Schaan}, E., {Ferraro}, S., {Vargas-Maga{\~n}a}, M., {et~al.} 2016, \prd, 93,
  082002

\bibitem[{{Scoccimarro} {et~al.}(2001){Scoccimarro}, {Sheth}, {Hui}, \&
  {Jain}}]{Scoccimarro2001}
{Scoccimarro}, R., {Sheth}, R.~K., {Hui}, L., \& {Jain}, B. 2001, \apj, 546, 20

\bibitem[{{Sembolini} {et~al.}(2013){Sembolini}, {Yepes}, {De Petris},
  {Gottl{\"o}ber}, {Lamagna}, \& {Comis}}]{Sembolini2013}
{Sembolini}, F., {Yepes}, G., {De Petris}, M., {et~al.} 2013, \mnras, 429, 323

\bibitem[{{Shull} {et~al.}(2012){Shull}, {Smith}, \& {Danforth}}]{shull2012}
{Shull}, J.~M., {Smith}, B.~D., \& {Danforth}, C.~W. 2012, \apj, 759, 23

\bibitem[{{Springel} {et~al.}(2005){Springel}, {White}, {Jenkins}, {Frenk},
  {Yoshida}, {Gao}, {Navarro}, {Thacker}, {Croton}, {Helly}, {Peacock}, {Cole},
  {Thomas}, {Couchman}, {Evrard}, {Colberg}, \& {Pearce}}]{Springel2005}
{Springel}, V., {White}, S.~D.~M., {Jenkins}, A., {et~al.} 2005, \nat, 435, 629

\bibitem[{{Sunyaev} \& {Zeldovich}(1972)}]{SZ1972}
{Sunyaev}, R.~A. \& {Zeldovich}, Y.~B. 1972, Comments on Astrophysics and Space
  Physics, 4, 173

\bibitem[{{Tanimura} {et~al.}(2019){Tanimura}, {Hinshaw}, {McCarthy}, {Van
  Waerbeke}, {Aghanim}, {Ma}, {Mead}, {Hojjati}, \&
  {Tr{\"o}ster}}]{tanimura2017}
{Tanimura}, H., {Hinshaw}, G., {McCarthy}, I.~G., {et~al.} 2019, \mnras, 483,
  223

\bibitem[{{Tejos} {et~al.}(2016){Tejos}, {Prochaska}, {Crighton}, {Morris},
  {Werk}, {Theuns}, {Padilla}, {Bielby}, \& {Finn}}]{Tejos2016}
{Tejos}, N., {Prochaska}, J.~X., {Crighton}, N.~H.~M., {et~al.} 2016, \mnras,
  455, 2662

\bibitem[{{Vikram} {et~al.}(2017){Vikram}, {Lidz}, \& {Jain}}]{Vikram2017}
{Vikram}, V., {Lidz}, A., \& {Jain}, B. 2017, \mnras, 467, 2315

\bibitem[{{White} {et~al.}(2011){White}, {Blanton}, {Bolton}, {Schlegel},
  {Tinker}, {Berlind}, {da Costa}, {Kazin}, {Lin}, {Maia}, {McBride},
  {Padmanabhan}, {Parejko}, {Percival}, {Prada}, {Ramos}, {Sheldon}, {de
  Simoni}, {Skibba}, {Thomas}, {Wake}, {Zehavi}, {Zheng}, {Nichol},
  {Schneider}, {Strauss}, {Weaver}, \& {Weinberg}}]{White2011}
{White}, M., {Blanton}, M., {Bolton}, A., {et~al.} 2011, \apj, 728, 126

\bibitem[{{Zheng} {et~al.}(2007){Zheng}, {Coil}, \& {Zehavi}}]{Zheng2007}
{Zheng}, Z., {Coil}, A.~L., \& {Zehavi}, I. 2007, \apj, 667, 760

\end{thebibliography}

\begin{appendix}

\section{Estimating the density with the cylindrical filament model}\label{sec:derivation}
From the observed SZ effect of the gas filament and the corresponding total matter of the filament from CMB lensing measurements, we can estimate both the gas and matter density.
We assume that the filament takes the shape of a cylinder and both the gas and total matter follow a two-dimensional Gaussian profile perpendicular to the direction of the filament. In the following we take the derivation for the gas component for clarity. It applies to the total matter density by simply changing the notation. The gas density perpendicular to the filament direction can be expressed as 
 \begin{equation}
 n_{\mathrm{e}}(\ell, r_{\perp})=n_{\mathrm{0}}\exp{\left[-\frac{r_{\perp}^2}{2(\sigma^2+\sigma_B^2)}\right]}\exp{\left(-\frac{\ell^2}{2\sigma^2}\right)}, 
 \end{equation}
 where $(\ell,r_{\perp})$ are coordinates in the plane perpendicular to the filament direction, within which ${\bf \ell}$ is the line-of-sight direction and ${\bf r_{\perp}}$ is the direction perpendicular to the line of sight. The intrinsic Gaussian width of the filament is denoted by $\sigma$, and $\sigma_\mathrm{B}$ is the corresponding width of the Gaussian \textit{Planck} beam of FWHM 10 arcmin. The density profile is convolved by the 10-arcmin FWHM Gaussian beam along ${\bf r_{\perp}}$. Using Eq.~\ref{eq:tsz}, the averaged one-dimensional $y$ profile along the ${\bf r_{\perp}}$ direction is
 \begin{equation}
 y(r_{\perp}) = 
 \sqrt{2\pi}\,n_{\rm 0}\,\sigma \frac{k_\mathrm{B} T_\mathrm{e}\sigma_\mathrm{T}}{m_\mathrm{e} c^2} \exp{\left[-\frac{r_{\perp}^2}{2(\sigma^2+\sigma_\mathrm{B}^2)}\right]}.
 \end{equation}
 We find from our measurement that the peak value of $y(r_{\perp})$ is very close to its mean within the box region of the filament (Fig.~1), i.e. $y_{\rm peak}\simeq \bar{y}/0.9$.
 This yields 
 \begin{equation}
 n_{\rm 0}=\frac{\bar y/0.9}{\sqrt{2\pi}\sigma} \frac{m_\mathrm{e} c^2}{k_\mathrm{B} T_\mathrm{e}\sigma_\mathrm{T}}.   
 \end{equation}
Therefore, given the measurement of $\bar y$ from data, the exact value of the central density of the filament depends on the assumption for the Gaussian $\sigma$ of the filament as expected. However, the total number of electrons in the filament 
\begin{equation}
N_{\mathrm{e}}=L\int n_{\mathrm{e}}(\ell,r_{\perp})\,d\ell\, dr_{\perp} =\frac{\bar y}{0.9} \frac{m_\mathrm{e} c^2}{k_\mathrm{B} T_\mathrm{e}\sigma_\mathrm{T}}   \sqrt{2\pi}\,L\,\sqrt{\sigma^2+\sigma^2_\mathrm{B}}
\end{equation}
where $L$ is the length of the filament, is insensitive to that assumption at the limit where $\sigma_\mathrm{B}\gg \sigma$, which is our case as the beam of the CMB map is dominant. Here, $\sigma_\mathrm{B} \simeq 3\times\sigma$, so the choice of $\sigma$ makes a difference of $\sim\,$1 percentage point for the baryon fraction estimate.\\\\
With the same derivation, the total matter density contrast at the centre of the filament $\delta_0$ is related to the mean lensing convergence $\bar \kappa$ via
\begin{equation}
\delta_{\rm 0}=\frac{\bar \kappa/0.9}{\sqrt{2\pi}\sigma} \,\frac{2a c^2}{3 H_0^2\Omega_{\rm m}}\,\frac{D_\mathrm{S}}{D_\mathrm{L}(D_\mathrm{S}-D_\mathrm{L})}.  
\end{equation}
Using the above equation with the observed value of $\bar \kappa$, the central matter density is estimated to be $\rho_{\rm 0}(z)=[\delta_{\rm 0}(z) +1]\bar \rho(z) \simeq 5.5 \times \bar{\rho}(z)$. Assuming that the baryon fraction in the region of the filament follows the mean value of the Universe, the central gas density of the filament is therefore also $n_{\rm 0}(z) \simeq 5.5 \times \bar n_{\rm e}(z)$, and so the corresponding gas temperature is approximately $2.7\times 10^6\,$K.

\end{appendix}

\end{document}